\documentclass[12pt]{article}
\usepackage{indent}
\usepackage{eqsection}
\usepackage{subeqnarray}
\usepackage{amsfonts}

\def\Z{{\mathbb Z}}
\def\R{{\mathbb R}}
\def\C{{\mathbb C}}
\def\Q{{\mathbb Q}}

\begin{document}

\bibliographystyle{plain}

\date{October 15, 2001 \\ Revised: December 17, 2001}

\title{\vspace*{-1cm} 
Exact Finite-Size-Scaling Corrections to the \\
       Critical Two-Dimensional Ising Model on a Torus. \\[5mm]
       \large\bf II.~Triangular and hexagonal lattices}

\author{
  {\small Jes\'us Salas}                                    \\[-0.2cm]
  {\small\it Departamento de F\'{\i}sica Te\'orica}         \\[-0.2cm]
  {\small\it Facultad de Ciencias, Universidad de Zaragoza} \\[-0.2cm]
  {\small\it Zaragoza 50009, SPAIN}                         \\[-0.2cm]
  {\small\tt JESUS@MELKWEG.UNIZAR.ES}                       \\[-0.2cm]
  {\protect\makebox[5in]{\quad}}  
  \\
}
\vspace{0.5cm}

\maketitle
\thispagestyle{empty}   

%
%

\def\spose#1{\hbox to 0pt{#1\hss}}
\def\ltapprox{\mathrel{\spose{\lower 3pt\hbox{$\mathchar"218$}}
 \raise 2.0pt\hbox{$\mathchar"13C$}}}
\def\gtapprox{\mathrel{\spose{\lower 3pt\hbox{$\mathchar"218$}}
 \raise 2.0pt\hbox{$\mathchar"13E$}}}
\def\inapprox{\mathrel{\spose{\lower 3pt\hbox{$\mathchar"218$}}
 \raise 2.0pt\hbox{$\mathchar"232$}}}

%
%


\begin{abstract}
We compute the finite-size corrections to the free energy, internal energy
and specific heat of the critical two-dimensional spin-$1/2$ Ising model
on a triangular and hexagonal lattices wrapped on a torus. We find the 
general form of the finite-size corrections to these quantities, as well as
explicit formulas for the first coefficients of each expansion. We analyze the 
implications of these findings on the renormalization-group description of 
the model. 
\end{abstract}



\clearpage

%
%
\newcommand{\be}{\begin{equation}}
\newcommand{\ee}{\end{equation}}
\newcommand{\<}{\langle}
\renewcommand{\>}{\rangle}
\newcommand{\para}{\|}
\renewcommand{\perp}{\bot}

\def\smfrac#1#2{{\textstyle\frac{#1}{#2}}}
\def\half{ {{1 \over 2 }}}
\def\smhalf{ {\smfrac{1}{2}} }
\def\scra{{\cal A}}
\def\scrc{{\cal C}}
\def\scrd{{\cal D}}
\def\scre{{\cal E}}
\def\scrf{{\cal F}}
\def\scrg{{\cal G}}
\def\scrh{{\cal H}}
\def\scrj{{\cal J}}
\def\scrk{{\cal K}}
\def\scrl{{\cal L}}
\def\scrm{{\cal M}}
\newcommand{\scrmvec}{\vec{\cal M}_V}
\def\scrmtens{{\stackrel{\leftrightarrow}{\cal M}_T}}
\def\scro{{\cal O}}
\def\scrp{{\cal P}}
\def\scrr{{\cal R}}
\def\scrs{{\cal S}}
\def\ttens{{\stackrel{\leftrightarrow}{T}}}
\def\scrv{{\cal V}}
\def\scrw{{\cal W}}
\def\scry{{\cal Y}}
\def\tauss{\tau_{int,\,\scrm^2}}
\def\taux{\tau_{int,\,{\cal M}^2}}
\newcommand{\taum}{\tau_{int,\,\vec{\cal M}}}
\def\taue{\tau_{int,\,{\cal E}}}
\newcommand{\imag}{\mathop{\rm Im}\nolimits}
\newcommand{\real}{\mathop{\rm Re}\nolimits}
\newcommand{\tr}{\mathop{\rm tr}\nolimits}
\newcommand{\sgn}{\mathop{\rm sgn}\nolimits}
\newcommand{\codim}{\mathop{\rm codim}\nolimits}
\newcommand{\rank}{\mathop{\rm rank}\nolimits}
\newcommand{\sech}{\mathop{\rm sech}\nolimits}
\def\textprime{{${}^\prime$}}
\newcommand{\longto}{\longrightarrow}
\def\var{ \hbox{var} }
\newcommand{\gtilde}{ {\widetilde{G}} }
\newcommand{\USp}{ \hbox{\it USp} }
\newcommand{\CP}{ \hbox{\it CP\/} }
\newcommand{\QP}{ \hbox{\it QP\/} }
\def\hboxscript#1{ {\hbox{\scriptsize\em #1}} }

\newcommand{\plotdot}{\makebox(0,0){$\bullet$}}
\newcommand{\plotsmalldot}{\makebox(0,0){{\footnotesize $\bullet$}}}

\def\bsigma{\mbox{\protect\boldmath $\sigma$}}
\def\bpi{\mbox{\protect\boldmath $\pi$}}
\def\btau{\mbox{\protect\boldmath $\tau$}}
\def\bn{{\bf n}}
\def\br{{\bf r}}
\def\bz{{\bf z}}
\def\bh{\mbox{\protect\boldmath $h$}}

\def\betatilde{ {\widetilde{\beta}} }
\def\hatp{\hat p}
\def\hatl{\hat l}

\def\msbar{ {\overline{\hbox{\scriptsize MS}}} }
\def\normalmsbar{ {\overline{\hbox{\normalsize MS}}} }

\def\eff{ {\hbox{\scriptsize\em eff}} }

\newcommand{\reff}[1]{(\ref{#1})}

\def\N{\hbox{$\mathbb N$}}
\def\C{\hbox{$\mathbb C$}}
\def\Q{\hbox{$\mathbb Q$}}
\def\R{\hbox{$\mathbb R$}}
\def\Z{\hbox{$\mathbb Z$}}

\newtheorem{theorem}{Theorem}[section]
\newtheorem{corollary}[theorem]{Corollary}
\newtheorem{lemma}[theorem]{Lemma}
\newtheorem{conjecture}[theorem]{Conjecture}
\newtheorem{definition}[theorem]{Definition}
\def\proof{\bigskip\par\noindent{\sc Proof.\ }}
\def\qed{\hbox{\hskip 6pt\vrule width6pt height7pt depth1pt \hskip1pt}\bigskip}

%
%
\newenvironment{sarray}{
          \textfont0=\scriptfont0
          \scriptfont0=\scriptscriptfont0
          \textfont1=\scriptfont1
          \scriptfont1=\scriptscriptfont1
          \textfont2=\scriptfont2
          \scriptfont2=\scriptscriptfont2
          \textfont3=\scriptfont3
          \scriptfont3=\scriptscriptfont3
        \renewcommand{\arraystretch}{0.7}
        \begin{array}{l}}{\end{array}}

\newenvironment{scarray}{
          \textfont0=\scriptfont0
          \scriptfont0=\scriptscriptfont0
          \textfont1=\scriptfont1
          \scriptfont1=\scriptscriptfont1
          \textfont2=\scriptfont2
          \scriptfont2=\scriptscriptfont2
          \textfont3=\scriptfont3
          \scriptfont3=\scriptscriptfont3
        \renewcommand{\arraystretch}{0.7}
        \begin{array}{c}}{\end{array}}
%
%
%
%

%
%
\section{Introduction}   \label{sec_intro}

It is well-known that phase transitions in statistical-mechanical systems 
can occur only in the infinite-volume limit.  In any finite system,
all thermodynamic quantities (such as the magnetic susceptibility
and the specific heat) are analytic functions of all parameters
(such as the temperature and the magnetic field);
but near a critical point they display peaks whose height increases
and whose width decreases as the volume $V=L^d$ grows,
yielding the critical singularities in the limit $L \to\infty$.
For bulk experimental systems (containing $V \sim 10^{23}$ particles)
the finite-size rounding of the phase transition is usually beyond
the experimental resolution;
but in Monte Carlo simulations ($V\ltapprox 10^6$--$10^7$)
it is visible and is often the dominant effect.

Finite-size scaling theory
\cite{Fisher,Fisher_Barber,Barber,Privman}
provides a systematic framework for understanding finite-size effects
near a critical point.
The idea is simple:  the only two relevant length scales are
the system linear size $L$ and the correlation length $\xi_\infty$ of the
bulk system at the same parameters, so everything is controlled by the
single ratio $\xi_\infty/L$.\footnote{
   This is true only for systems below the upper critical dimension $d_c$. 
   For Ising models with short-range interaction, $d_c = 4$.
}
If $L \gg \xi_\infty$, then finite-size effects are negligible;
for $L \sim \xi_\infty$, thermodynamic singularities are rounded
and obey a scaling Ansatz
${\cal O} \sim L^{p_{\cal O}} F_{\cal O}(\xi_\infty/L)$
where $p_{\cal O}$ is a critical exponent and
$F_{\cal O}$ is a scaling function.
Finite-size scaling is the basis of the powerful 
phenomenological renormalization group method
(see ref.~\cite{Barber} for a review);
and it is an efficient tool for extrapolating finite-size data
coming from Monte Carlo simulations so as to obtain accurate results on
critical exponents, universal amplitude ratios and subleading exponents
\cite[and references therein]{fss_greedy,MGMC_SU3,%
Salas_Sokal_Ising_published,XY}.\footnote{
  Finite-size scaling has also been successfully applied to data coming 
  from transfer-matrix computations \cite{Blote}.
} 
In particular, in systems with multiplicative and/or additive logarithmic
corrections (as the two-dimensional 4-state Potts model \cite{Salas_Sokal_FSS}),
a good understanding of finite-size effects is crucial for obtaining reliable
estimates of the physically interesting quantities.

In finite-size-scaling theory for systems with periodic boundary conditions,
three simplifying assumptions have frequently been made:
\begin{itemize}
\item[(a)] The regular part of the free energy, $f_{\rm reg}$,
           is independent of the lattice size $L$ \cite{Privman}
           (except possibly for terms that are exponentially small in $L$).
\item[(b)] The scaling fields associated to the temperature $T$ and magnetic 
           field $h$ (i.e., $\mu_t$ and $\mu_h$, respectively) are independent 
           of $L$ \cite{Guo_87}.
\item[(c)] The scaling field $\mu_L$ associated to the lattice size 
           equals $L^{-1}$ exactly, with no corrections $L^{-2}$,
           $L^{-3}$, \ldots\  \cite{Privman}.
\end{itemize}
Moreover, in the nearest-neighbor spin-1/2 two-dimensional Ising model,
it was further assumed for many years that there are no irrelevant operators 
\cite{Aharony_83,Gartenhaus_88}; indeed this assumption was confirmed 
numerically through order $(T-T_c)^3$ at least as regards the bulk 
behavior of the susceptibility in the isotropic square-lattice Ising model
\cite{Gartenhaus_88}. However, 
several authors have recently found overwhelming evidence that there are 
indeed irrelevant operators playing a role in the two-dimensional Ising model 
\cite{Nickel_a,Nickel_b,Pelissetto_00,Orrick_00a,Orrick_00b,%
Caselle_01a,Caselle_01b}. 
In particular, for the square-lattice Ising model they have
found by studying the bulk magnetic susceptibility that there is one 
irrelevant operator contributing to order $(T-T_c)^4$ and there is (at 
least) one irrelevant operator contributing to order $(T-T_c)^6$. 

An interesting way to test assumptions (a)--(c) and see the effect of 
the irrelevant operators is to compute the asymptotic expansion 
(in powers of $L^{-1}$) of the free energy and its derivatives with respect
to the temperature at the critical point. The square-lattice Ising model
is the best understood case.  

In a classic paper, Ferdinand and Fisher \cite{Ferdinand_Fisher} 
considered the energy and the specific heat of the square-lattice Ising model
on a torus of width $L$ and aspect ratio $\rho$,
and obtained the first two (resp. three) terms
of the large-$L$ asymptotic expansion of the energy (resp.\ specific heat)
at fixed $x \equiv L(T-T_c)$ [this is the finite-size-scaling regime]
and fixed $\rho$. In particular, at criticality ($T=T_c$) 
they computed the finite-size corrections to both quantities to order $L^{-1}$. 
Their results have been improved at the critical point by several 
authors \cite{Izmailian,Izmailian2,Salas_01,Izmailian4}. Their results 
can be summarized as follows:
\begin{subeqnarray}
\slabel{results_square_f}
f_c^{\rm sq}(L,\rho) &=& f_{\rm bulk}^{\rm sq} + 
                    \sum\limits_{m=1}^\infty 
                 {f_{2m}^{\rm sq}(\rho)\over L^{2m}} \\
E_c^{\rm sq}(L,\rho) &=& E_0 + \sum\limits_{m=0}^\infty 
            {E_{2m+1}^{\rm sq}(\rho)\over L^{2m+1}} \\
C_{H,c}^{\rm sq}(L,\rho) &=& C_{00}^{\rm sq}\log L + C_0^{\rm sq}(\rho) + 
              \sum\limits_{m=1}^\infty {C_m^{\rm sq}(\rho) \over L^m}
\label{results_square}
\end{subeqnarray}
where $f_c$, $E_c$ and $C_{H,c}$ are respectively the critical free energy, 
internal energy and specific heat.\footnote{Janke and Kenna 
\protect\cite{Janke_01} have studied
similar expansions for the square-lattice Ising model with Branscamp-Kunz
boundary conditions. The analytic structure is similar to 
\protect\reff{results_square} but additional terms arise due to the 
boundary conditions. For instance, there is a term $\sim \log L/L$ in the 
specific heat. On the other hand, Lu and Wu \cite{Lu} studied the 
critical free energy for the square-lattice Ising model on non-orientable 
surfaces (namely, the M\"obius strip and the Klein bottle). 
They found the first terms of the large-$L$ expansion
of $f_c(L,\rho)$; although they did not give details about the analytic 
structure of such expansion. In particular, there is an additional term 
$\sim L^{-1}$ in the expansion for the M\"obius strip (due to ``surface'' 
effects) which is absent in the Klein bottle. They also explicitly showed 
that the coefficient $f_2^{\rm sq}(\rho)$ depends on the boundary conditions 
[even if the expansion \protect\reff{results_square_f} holds true].
} 

The first important observation is that there are no logarithmic corrections
except for the specific-heat leading term $C_{00}^{\rm sq}\log L$. Secondly, 
the finite-size corrections are integer powers of $L^{-1}$, which is consistent
with irrelevant operators taking integer exponents. Furthermore, not all
the powers of $L^{-1}$ occur: in the large-$L$ expansion of the 
free energy (resp.\ internal energy) only even (resp.\ odd) powers of 
$L^{-1}$ can occur. In the specific-heat expansion all powers of $L^{-1}$ 
can appear. In addition, the coefficients $C_m^{\rm sq}$ and 
$E_m^{\rm sq}$ satisfy the relation 
\be
\label{ratio_sq}
{E_{m}^{\rm sq}(\rho) \over C_{m}^{\rm sq}(\rho) } = \left\{ \begin{array}{ll}
                         - 1/\sqrt{2} & \qquad \hbox{\rm for $m$ odd} \\ 
                           0          & \qquad \hbox{\rm for $m$ even} 
                                   \end{array} \right. 
\ee

The authors of ref.~\cite{Caselle_01b} classified (using conformal field 
theory) all possible irrelevant operators that may occur in the two-dimensional
Ising model and found that all their
results (in the thermodynamic limit and at criticality on a finite torus)
can be explained in terms of the following conjecture

\begin{conjecture} 
\label{conjecture_Caselle}
{\bf \protect\cite[Conjecture (d2)]{Caselle_01b}}
The only irrelevant operators which appear in the two-dimensional 
nearest-neighbor Ising model are those due to the lattice breaking of the
rotational symmetry
\end{conjecture}

In particular, for the square-lattice Ising model the first operator that
breaks rotational invariance is the spin-four operator 
$T^2 + \bar{T}^2$ (where here $T$ is the energy-momentum operator) whose 
renormalization-group exponent is $y=-2$. In ref.~\cite{Caselle_01b} they 
showed that this operator can give rise to all the observed corrections in 
\reff{results_square}.\footnote{
  A similar finite-size scaling analysis was carried out for the 
  one-dimensional Ising quantum chain which belongs to the same 
  universality class of the two-dimensional Ising model 
  \protect\cite{Henkel_87,Reinicke_87a,Reinicke_87b,Henkel_99}.
} 

In this paper we extend the above results to the 
triangular and hexagonal lattices. We will obtain the large-$L$ 
asymptotic expansions for the critical free energy, internal energy and
specific heat for such lattices wrapped on a torus of width $L$ and 
fixed aspect ratio $\rho$. The interest of this computation is 
three-fold. First, we can make a new test of 
Conjecture~\ref{conjecture_Caselle}. In the triangular lattice, the 
first irrelevant operator (belonging to the identity family) 
that breaks rotational invariance is $T^3 + \bar{T}^3$ with $y=-6$  
\cite{Caselle_01b}. If Conjecture~\ref{conjecture_Caselle} is true, then  
several coefficients in the finite-size-scaling expansions 
\reff{results_square} should vanish.  
Second, we can directly check whether the ratio \reff{ratio_sq} is 
universal or not, that is, if \reff{ratio_sq} depends or not on the 
microscopic details of the lattice. Finally, the asymptotic expansions
could be useful to check Monte Carlo simulations.

A first study of the triangular-lattice Ising model partition function 
on a finite torus was done by Nash and O'Connor \cite{Nash_99}. 
They obtained (among other
interesting results) the exact expression of such partition function with 
anisotropic nearest-neighbor couplings and extracted its scaling limit.  
They computed the bulk contribution to the free energy $f_{\rm bulk}$ 
and the first finite-size correction $f_2(\rho)$. Here we will extend their
results at the critical point.  

The main results of this paper can be summarized as follows: 
\begin{subeqnarray}
f_c(L,\rho) &=& f_{\rm bulk} + \sum\limits_{m=1}^\infty {f_{2m}(\rho)\over 
              L^{2m}} \\
E_c(L,\rho) &=& E_0 + \sum\limits_{m=0}^\infty {E_{2m+1}(\rho)\over 
              L^{2m+1}} \\ 
C_{H,c}(L,\rho) &=& C_{00}\log L + C_0(\rho) + \sum\limits_{m=1}^\infty
             {C_m(\rho) \over L^m} \\
f^{(3)}_c(L,\rho) &=& {\cal A}_1(\rho) L + A_{00}\log L + A_0(\rho) + 
                         \sum\limits_{m=1}^\infty {A_m(\rho) \over L^m} \\
f^{(4)}_{c,{\rm log}}(L,\rho) &=& B_{00}\log L 
\end{subeqnarray} 
where $f^{(3)}_c$ (resp.\  $f^{(4)}_{c,{\rm log}}$) is the third derivative 
(resp.\ the logarithmic contribution to the fourth derivative) of the free 
energy with respect to the inverse temperature $\beta$ evaluated at the 
critical point.
We have also found explicit formulas for the coefficients $f_2(\rho)$, 
$f_6(\rho)$, $E_1(\rho)$, $E_5(\rho)$, $C_{00}$, $C_0(\rho)$, $C_1(\rho)$, 
$C_4(\rho)$, $C_5(\rho)$, ${\cal A}_1(\rho)$,  $A_{00}$, $A_0(\rho)$, 
$A_1(\rho)$, and $B_{00}$ 
(Indeed, $f_4 = f_8 = E_3 = E_7 = C_2 = C_3 = A_2 = 0$).

Our results on the general analytic structure of the finite-size corrections
to these models are:
\begin{itemize}

\item The analytic structure of the finite-size-scaling corrections of the
quantities considered here is exactly the same for the triangular and the
hexagonal lattices.  

\item The finite-size corrections to the free energy, internal energy and 
specific heat are always integer powers of $L^{-1}$, {\em unmodified by 
logarithms}\/ (except of course for the leading $\log L$ term in the 
specific heat).

\item In the finite-size expansion of the free energy, only {\em even}\/ 
integer powers of $L^{-1}$ occur. The only exceptions are the powers 
$L^{-4}$ and $L^{-8}$ whose coefficients vanish.

\item In the finite-size expansion of the energy, we only find {\em odd}\/ 
integer powers of $L^{-1}$. In this case, the coefficients associated to the 
powers $L^{-3}$ and $L^{-7}$ vanish.  

\item In the finite-size expansion of the specific heat, any integer powers
of $L^{-1}$ can occur, except the terms $L^{-2}$ and $L^{-3}$.  
In addition, the non-zero coefficients of the odd powers 
of $L^{-1}$ in this expansion are proportional to the corresponding 
coefficients in the internal energy expansion as in the square lattice.    

\item In the finite-size expansion of $f^{(3)}_c$ we find that the 
expected leading term $L\log L$ is missing, and the actual leading term is 
simply $L$. We find that all powers of $L^{-1}$ appear in such expansion, 
except $L^{-2}$.  

\item In the finite-size expansion of the fourth derivative of the free 
energy $f^{(4)}_c$ we find that there is only a logarithmic term 
$\sim\log L$, even though we expect two additional logarithmic contributions
of order $L \log L$ and $L^2 \log L$ respectively.  

\end{itemize}

The above results are very useful to gain new insights on the 
renormalization-group description of the two-dimensional Ising model. 
Our conclusions on this topic are

\begin{itemize}

\item Some irrelevant operators should vanish at criticality. 
      This happens, in particular to the less irrelevant one 
      $T\bar{T}$ with renormalization-group exponent $y=-2$. 

\item In order to give account of all the finite-size corrections, we should 
      include {\em at least} two irrelevant operators, in agreement with the 
      results of \cite{Orrick_00a,Orrick_00b}. 

\item The scaling function $\widehat{W}(x)$ (which is the responsible for the 
      logarithmic corrections to the derivatives of the free energy) 
      vanishes at criticality $x=0$. Its first derivatives at criticality
      satisfy 
\be
\left.{\partial^n W(x) \over \partial x^n}\right|_{x=0} = 
  \left\{ \begin{array}{ll}
           0  & \qquad \hbox{\rm for $n=1,3,4$} \\
           1/(\lambda \pi\sqrt{3}) & \qquad \hbox{\rm for $n=2$} 
           \end{array}
  \right.
\ee
      where $\lambda= 1$ (resp.\   2) for the triangular (resp.\  hexagonal)
      lattice. These equations motivate the conjecture that 
      $\widehat{W}(x) = x^2/(2\lambda\pi\sqrt{3})$. 

\item The non-linear scaling field associated to the temperature can  
      be computed for both lattices and it is given by  
\be
      \mu_t(\tau) = \tau - {1 \over 24}\tau^3 + {\cal O}(\tau^5)
\ee
     This result provides a cross-check of the analysis of infinite-volume 
     quantities \cite{Caselle_01b}.

\end{itemize}

The plan of this paper is as follows:
In Section~\ref{sec_definitions} we present our definitions and notation.  
In Sections~\ref{sec_F}, \ref{sec_E}, \ref{sec_CH} and \ref{sec_der_CH} 
we present the computation of the 
asymptotic expansions for the free energy, internal energy, specific heat, and
higher derivatives of the free energy, respectively. 
In Section~\ref{sec_irrel} we discuss the  
consequences of our results on the renormalization-group description of the
models. In particular, we will focus on the irrelevant operators of the
model and on the finite-size-scaling functions.  
Finally, in Section~\ref{sec_conclusions} we present our 
conclusions and discuss the results. 
We have summarized the technical details in the appendixes: in 
Appendix~\ref{sec_Euler} we recall the Euler-MacLaurin formula, and in 
Appendix~\ref{sec_theta} (resp.\  Appendix~\ref{sec_Kronecker}) we collect 
the definitions and properties of the $\theta$-functions (resp.\ Kronecker's
double series).  

%
%
\section{Basic definitions} \label{sec_definitions}

Let us first consider an Ising model on a triangular lattice wrapped 
on a torus of size $N \times M$ at zero magnetic field. The Hamiltonian 
is given by 
\be
\label{Ising_Hamiltonian}
{\cal H} = - \beta \sum\limits_{<i,j>}\sigma_i \sigma_j
\ee
The partition function is given by 
\be
\label{Ising_partition_function}
Z_{N M}(\beta) = \sum\limits_{\{\sigma=\pm1\}} e^{-{\cal H}} 
\ee
The dual of such triangular lattice is an hexagonal lattice wrapped on a 
torus of size $N\times M$ and containing $2 N M$ sites (i.e., the hexagonal
lattice can be viewed as a triangular lattice with a two-point basis). 
The Hamiltonian and the partition function of the Ising model on this lattice
are also given by \reff{Ising_Hamiltonian}/\reff{Ising_partition_function}.

If one brushes aside some subtleties about boundary conditions,
one can relate the partition function \reff{Ising_partition_function}
of a triangular-lattice Ising model
at coupling $\beta$ to the partition function
of the Ising model on the dual (i.e., hexagonal) lattice at a ``dual'' coupling
$\beta^\star$ \cite{Savit_80,Wu_82}:
\be
\label{def_duality}
Z_{NM}^{\rm tri}(\beta) = Z_{2NM}^{\rm hc}(\beta^\star) \, 2^{1-2NM}\,  
   (2 \sinh 2\beta)^{3NM/2}
\ee
where $\beta^\star$ is defined by  
\be
\label{def_beta_star} 
\tanh \beta^\star = e^{-2\beta} 
\ee
Using eq.~\reff{def_duality} and the star-triangle equation 
\cite{Baxter_book} we can obtain the critical values of the couplings for
both models
\be
\label{def_beta_c}
\beta_c = \left\{ \begin{array}{ll}
             {1\over 4}\log 3             & \qquad \qquad {\rm triangular}
             \\[2mm] 
             {1\over 2}\log (2 +\sqrt{3}) & \qquad \qquad {\rm hexagonal}
                \end{array}\right.
\ee
However, this argument is strictly valid only in the infinite-volume limit;
it gives the correct relation
\be
   f^{\rm tri}(\beta) = 2 f^{\rm hc}(\beta^\star) - 2\log 2 + 
   {3\over 2} \log( 2\sinh 2\beta ) 
\ee
between infinite-volume free energies and the correct critical points 
\reff{def_beta_c}, but the identity \reff{def_duality} for finite-lattice 
partition functions does {\em not}\/ in general hold.
This is because a periodic lattice is non-planar,
so that the correct duality formula also involves a pair of ``homological''
modes arising from the two directions of winding around the torus 
\cite{duality}. 
Or put it another way:  high-temperature graphs that wind around the lattice
do {\em not}\/ necessarily correspond to low-temperature graphs
on the dual lattice. Therefore, on a finite lattice 
--- which is the subject of this paper  --- we need to be more 
careful.\footnote{
  We thank Alan Sokal for useful clarifications about this point.
}

We begin by computing the exact partition function of both models 
on a torus of size $N\times M$. One way to do this is by relating 
the Ising model to a dimer model \cite{McCoy}. The same computation leading to 
the square-lattice partition function can be used to obtain the 
hexagonal-lattice partition function \cite{Montroll} by changing the 
weights of the different dimer configurations. 
Though the triangular-lattice Ising
partition function cannot be derived from the hexagonal-lattice
partition function using duality \reff{def_duality},
for the reasons given above,
we can instead use the star-triangle transformation \cite{Baxter_book}.
Then, the triangular-lattice partition function $Z_{MN}^{\rm tri}$ is 
related to the hexagonal-lattice partition function 
$Z_{2MN}^{\rm hc}$ (containing twice as much sites) by the formula 
\be
Z_{MN}^{\rm tri}(\beta) = R(\beta)^{-MN} Z_{2MN}^{\rm hc}(\tilde{\beta}) 
\label{def_star_triangle}
\ee
where the $\tilde{\beta}$ and $R(\beta)$ are given by \cite{Baxter_book}
\begin{subeqnarray}
\sinh 2 \tilde{\beta} &=& {1\over \kappa(\beta)} {1\over \sin 2\beta} \\
R(\beta)^2            &=& {2 \over \kappa(\beta)^2 \sinh^3 2\beta}  
\end{subeqnarray} 
and $\kappa$ (which depends on $\beta$ through the parameter $v=\tanh \beta$) 
is equal to \cite{Baxter_book}
\be
\kappa(\beta) = { (1-v^2)^3 \over 4 \sqrt{ (1+v^3)v^3(1+v)^3 }} 
\ee
After straightforward (but lengthy) algebra we find that the partition 
function for both lattices can be written in a very similar way 
in the ferromagnetic regime:
\be
\label{partition_function}
Z_{V}(\beta) = {1\over 2} (2 \sin 2\beta)^{V/2} \sum_{\alpha,\beta=0,1/2}
               Z_{\alpha,\beta}(\mu) 
\ee
where $V$ is the number of spins in the lattice (e.g., $V=N M$ in the 
triangular lattice and $V=2N M$ in the hexagonal lattice). The functions
$Z_{\alpha,\beta}(\mu)$ are given by
\begin{eqnarray}
Z_{\alpha,\beta}(\mu)^2 &=& \prod\limits_{n=0}^{N-1} \prod\limits_{m=0}^{M-1} 
 4 \left\{ \sin^2 \left( {\pi (n+\alpha)\over N}\right) + 
           \sin^2 \left( {\pi (m+\beta) \over M} \right) \right. \nonumber \\
                        & & \qquad \qquad \qquad + \left. 
           \sin^2 \left( {\pi (m+\beta) \over M} - 
                         {\pi (n+\alpha)\over N} \right) +2 \sinh^2 \mu 
  \right\} 
\label{def_Zabsquare}
\end{eqnarray}
where the ``mass'' term $\mu$ is given by
\be
e^{2 \mu}  = \left\{ \begin{array}{ll} 
                       {1\over 2}\left( e^{4\beta} -1\right) 
                                       & \qquad \qquad {\rm triangular}\\[2mm]
                       2 \sinh^2 \beta & \qquad \qquad {\rm hexagonal} 
                      \end{array}
                \right.
\label{def_mass}
\ee
The critical point corresponds to the vanishing of the mass, thus giving 
\reff{def_beta_c}.  

\bigskip

\noindent
{\bf Remark}. 
The fact that the partition function of both lattices depends on 
the same functions $Z_{\alpha,\beta}(\mu)$ can be explained by noting that the
translational symmetry of both lattices is the same (i.e., they have the same
underlying Bravais lattice). This issue explains why the finite-size 
expansions are so similar in both lattices.

\bigskip

The functions $Z_{\alpha,\beta}(\mu)$ can be expanded in powers of $\mu$.
In particular, when $(\alpha,\beta) \neq (0,0)$ the functions are even in 
$\mu$, while $Z_{0,0}(\mu)$ is an odd function of $\mu$: 
\begin{eqnarray}
\label{properties_Zab} 
Z_{\alpha,\beta}(\mu) &=& Z_{\alpha,\beta}(0) + {1\over 2!} 
                          Z_{\alpha,\beta}''(0) \mu^2 + \cdots \qquad 
                          (\alpha,\beta) \neq (0,0) \\ 
Z_{0,0}(\mu)          &=& \mu Z'_{\alpha,\beta}(0) + {1\over 3!} 
                          Z_{\alpha,\beta}'''(0) \mu^3 + \cdots
\end{eqnarray}
This is similar to what happens in the square-lattice Ising model 
\cite{Izmailian4}. 

We are interested in computing the asymptotic expansions for large $N$ and $M$
with fixed aspect ratio (e.g. length to width ratio):  
\be
\label{def_rho}
\rho = {M\over N}
\ee 
of the free energy $f(\beta;N,\rho)$, internal energy $E(\beta;N,\rho)$ and
specific heat $C_H(\beta;N,\rho)$ at the critical point $\beta=\beta_c$.
These quantities are defined as follows
\begin{subeqnarray}
\slabel{def_free_energy}
f(\beta;N,\rho)   &=& {1\over V} \log Z_{V}(\beta) \\  
\slabel{def_energy}
E(\beta;N,\rho)   &=& - {\partial \over \partial \beta} f(\beta;N,\rho) \\
\slabel{def_specific_heat}
C_H(\beta;N,\rho) &=&   {\partial^2 \over \partial \beta^2} f(\beta;N,\rho)
\end{subeqnarray}
In Section~\ref{sec_der_CH} we will also consider higher derivatives 
of the free energy at criticality 
\be
f^{(k)}_c(N,\rho)  = \left. {\partial^k  \over 
              \partial\beta^k} f(\beta;N,\rho) \right|_{\beta=\beta_c}  
\label{def_der_CH}
\ee
with $k=3,4$.

\bigskip

\noindent
{\bf Remark}. The definition of the specific heat \reff{def_specific_heat} 
is somewhat non-standard as it does not contain the factor $\beta^2$. 

\bigskip

%
%
\section{Finite-size-scaling corrections to the free energy}
\label{sec_F}

Let us start with the the basic quantity  $Z_{\alpha,\beta}$ 
\reff{def_Zabsquare} and write it in the form 
\begin{eqnarray}
Z_{\alpha,\beta}(\mu) &=&  \prod\limits_{n=0}^{N-1} \prod\limits_{m=0}^{M-1}
 4 \left\{ \cosh 2\mu + \sin^2 \left( {\pi (n+\alpha)\over N}\right) 
           \right. \nonumber \\
                        & & \qquad \qquad - \left.
           \cos \left( {\pi (n+\alpha)\over N} \right) 
           \cos \left( {2 \pi (m+\beta)\over M} -
                       {  \pi (n+\alpha)\over N} \right) 
  \right\}
\label{def_ZabsquareBis}
\end{eqnarray}
The product over $m$ in \reff{def_ZabsquareBis} can be exactly performed with
the help of the following identity \cite{Nash_99}: 
\be
\prod\limits_{m=0}^{M-1} \left[ \zeta - \lambda \cos \left( 
                      2\pi (m+\beta)\over M \right) \right] = 
\left( {\lambda z_{+}\over 2}\right)^M \left| 1 - z_{-} e^{- 2\pi i \beta} 
      \right|^2 
\label{formula_Nash}
\ee
where $\zeta$ and $\lambda$ are any two real numbers such that 
$|\zeta/\lambda| \geq 1$ and the quantities $z_\pm$ are given by 
\begin{subeqnarray}
\slabel{def_zpm}
z_{\pm} &=& {\zeta\over\lambda} \pm \sqrt{\left({\zeta\over\lambda}
                                          \right)^2 -1 } \\
z_{+} z_{-} &=& 1   
\slabel{property_zpm}
\end{subeqnarray}
We can finally write $Z_{\alpha,\beta}(\mu)$ as
\begin{eqnarray}
Z_{\alpha,\beta}(\mu) &=& 2^{N M/2} 
              \prod\limits_{n=0}^{N-1} 
              \left( \cosh 2\mu + \sin^2 \phi_{n+\alpha} 
              \phantom{ \sqrt{\left[\sin^2\right]^2}} 
                    \right. \nonumber \\
                      & & \qquad \qquad \qquad \left. 
                  +  \sqrt{ 
                     \left[ \cosh 2\mu + \sin^2 \phi_{n+\alpha} \right]^2 
                   - \cos^2 \phi_{n+\alpha} } 
              \right)^{M/2} 
    \nonumber \\
      & & \qquad \times  
              \prod\limits_{n=0}^{N-1} 
        \left| 
    1 - z_{-}(n+\alpha,N,\mu)^M  
        e^{-2\pi i \beta + M i \phi_{n+\alpha}} \right| 
\label{final_Zab}
\end{eqnarray}
where we have used the shorthand notations  
\begin{subeqnarray}
z_{\pm}(k,N,\mu) &=& 
          {\cosh 2\mu + \sin^2 \phi_k  \pm  
           \sqrt{ \left[ \cosh 2\mu + \sin^2 \phi_k \right]^2
                  - \cos^2 \phi_k } \over \cos \phi_k } \\ 
\phi_{k}         &=& {\pi k \over N}
\label{def_zpm_bis} 
\end{subeqnarray} 

Let us now evaluate the functions $Z_{\alpha,\beta}(0)$ for 
$(\alpha,\beta)\neq(0,0)$. We follow here the procedure used in 
ref.~\cite{Izmailian4}, which proved to be very efficient for extracting the
large-$N$ asymptotic expansions of the quantities of interest. 
We first compute the sum
\be
f_1 = {M\over 2}\sum\limits_{n=0}^{N-1} \log \left[ 
        1 + \sin^2 \phi_{n+\alpha} + \sin \phi_{n+\alpha} 
               \sqrt{3 + \sin^2 \phi_{n+\alpha}} \right] =
      {M\over 2}\sum\limits_{n=0}^{N-1} \omega_1(\phi_{n+\alpha}) 
\label{def_f1}
\ee
where 
\be
\label{def_w1}
\omega_1(k) = \log \left[ 1 + \sin^2 k + \sin k \sqrt{3 + \sin^2 k} \right]
            = \lambda k + \sum\limits_{k=2}^\infty {k^p \over p!} \lambda_p 
\ee 
The function $\omega_1$ and all its derivatives are integrable over 
$[0,\pi]$, and in addition, 
\be
\omega_1^{(k)}(\pi) - \omega_1^{(0)}(0) = \left\{ 
         \begin{array}{ll}
           -2 \omega_1^{(k)}(0) & \qquad k=2,6,10,12,14,\ldots \\
            0                   & \qquad {\rm otherwise} 
         \end{array} \right.
\ee
We can now use the Euler-MacLaurin summation formula 
\reff{Euler_MacLaurin_formula_final} to obtain 
\begin{eqnarray}
{1\over N}\sum\limits_{n=0}^{N-1} \omega_1(\phi_{n+\alpha}) &=& 
{1\over\pi} \int_0^\pi \omega_1(x) dx  
   - {\lambda\over \pi N^2} B_2(\alpha) \nonumber \\
   & & \qquad -  
   \sum\limits_{m=1}^\infty 
   \left( {\pi \over N}\right)^{2m} 
   {B_{2m+2}(\alpha) \over (2m+2)!} \lambda_{2m+1}
\label{paso_1}
\end{eqnarray}
The first coefficients $\lambda_k$ are 
\be
\label{relevant_l}
\lambda   = \sqrt{3}            \,; \quad 
\lambda_3 = \lambda_7 = 0       \,; \quad
\lambda_5 = {16\over \sqrt{3}}  \,; \quad  
\lambda_9 = 1792 \sqrt{3}       \,; \quad
\lambda_{11} 
          = - {51200\over \sqrt{3}}  
\ee
The final result for $f_1$ is  
\be
f_1 = {N M \over 2\pi} \int_0^\pi \omega_1(x) dx - 
      {\pi \lambda \rho \over 2} B_2(\alpha) - 
      \pi \rho \sum\limits_{m=1}^\infty 
      \left( {\pi \over N} \right)^{2m} 
      {B_{2m+2}(\alpha) \over (2m+2)!} \lambda_{2m+1} 
\label{result_f1}
\ee

Let us now consider the quantity $f_2$
\be
\label{def_f2}
f_2 = \sum\limits_{n=0}^{N-1} \log\left| 1 - z_{-}(n+\alpha,N,0)^M 
      e^{-2\pi i \beta + M i \phi_{n+\alpha} } \right| 
\ee
We first note that when $n+\alpha=N/2$, the factor 
$z_{-}(n+\alpha,N,0)=0$, so this term does not contribute to the sum 
\reff{def_f2}. In the other cases $z_{-}(n+\alpha,N,0)$ does not vanish
and we can use \reff{property_zpm} to write \reff{def_f2} as
\be
\label{def_f2_bis}
f_2 = \mathop{\textstyle \sum^{\, \prime}}_{n=0}^{N-1} \log\left| 1 -  
      e^{-M \log z_{+}(n+\alpha,N,0) -2\pi i \beta + M i \phi_{n+\alpha} } 
      \right|
\ee
where $\sum^\prime$ means that we have taken out the term with 
$n+\alpha=N/2$ (if such term exists). 

We now proceed as in ref.~\cite{Izmailian4}: we first write 
$\log | 1 - e^{-A}| = \real \log (1- e^{-A})$ and then we  
expand $\log (1- e^{-A})$ as a power series in $e^{-A}$: 
\be
\label{def_f2_tris}
f_2 = - \real \sum\limits_{p=1}^\infty 
              \mathop{\textstyle \sum^{\, \prime}}_{n=0}^{N-1} 
       {1\over p}  
      e^{-2p [M(\log z_+(n+\alpha,N,0) - i \phi_{n+\alpha})/2 +\pi i \beta] } 
\ee
It is convenient to write the function $\log z_{+}(k,N,0)$ as  
\be
\label{def_w2}
\log z_{+}(k,N,0) \equiv \omega_2(\phi_{k})  
             = \omega_1(\phi_{k}) - \log \cos \phi_{k} 
\ee
where $\omega_1(k)$ is the function \reff{def_w1}. We then split the sum 
over $n$ into two parts: $n \in [0, \lfloor N/2 \rfloor-1]$, and 
$n\in [\lfloor N/2 \rfloor,N-1]$. By making the substitution 
$n \rightarrow N-1-n$ in the second sum, we finally obtain  
\begin{eqnarray}
f_2 &=& - \real 
            \sum\limits_{p=1}^\infty 
            \mathop{\textstyle \sum^{\, \prime}}_{n=0}^{\lfloor N/2\rfloor-1} 
              {1\over p} 
              e^{-2p( M[w_2(\phi_{n+\alpha})- i\phi_{n+\alpha}] +i\pi \beta)}
          \nonumber \\
   & & \quad 
- \real 
           \sum\limits_{p=1}^\infty 
           \mathop{\textstyle \sum^{\, \prime}}_{n=0}^{N-\lfloor N/2\rfloor-1}
              {1\over p}
              e^{-2p( M[w_2(\phi_{n+1-\alpha})- i\phi_{n+\alpha}] -i\pi \beta)}
\end{eqnarray}

We now expand the function $\omega_2(k)$ as a power series in $k$
\be
\label{expansion_w2}
\omega_2(k) = \lambda k + \sum\limits_{m=1}^\infty 
              {\lambda_{2m+1}\over (2m+1)!} k^{2m+1} 
\ee
where the $\lambda_k$ are exactly those of the function $\omega_1$ 
\reff{relevant_l}. We obtain an expression of the form 
\begin{eqnarray}
f_2 &=& -\real \sum\limits_{p=0}^\infty {1\over p} 
               \mathop{\textstyle \sum^{\, \prime}}_{n=0}^{\lfloor N/2\rfloor-1}
               e^{-2p[\pi \tau_0 \rho (n+\alpha) + i \pi \beta]} 
               \nonumber \\
    & & \quad \quad \times \exp\left\{
          -\pi p \rho \sum\limits_{m=1}^\infty \left({\pi\over N}\right)^{2m} 
               {\lambda_{2m+1}\over (2m+1)! } (n+\alpha)^{2m+1} 
              \right\} \nonumber \\
    & & \quad -\real \sum\limits_{p=0}^\infty {1\over p} 
           \mathop{\textstyle \sum^{\, \prime}}_{n=0}^{N-\lfloor N/2\rfloor-1}
               e^{-2p[\pi \tau_0 \rho (n+1+\alpha) - i \pi \beta]} 
               \nonumber \\
    & & \quad \quad \quad \times \exp\left\{
          -\pi p \rho \sum\limits_{m=1}^\infty \left({\pi\over N}\right)^{2m}
               {\lambda_{2m+1}\over (2m+1)! } (n+1-\alpha)^{2m+1}
              \right\}
\label{def_f2_4}
\end{eqnarray}
where $\tau_0$ is a complex number equal to
\be
\label{def_tau0}
\tau_0 = {\lambda - i \over 2} = {\sqrt{3} - i \over 2} = e^{-i \pi/6}  
\ee

The next step consists in expanding the exponentials in powers of 
$N^{-k}$. By following the procedure introduced in 
\cite[Appendix B]{Izmailian4} we obtain  
\begin{eqnarray}
f_2 &=& -\real \sum\limits_{p=0}^\infty {1\over p}
               \mathop{\textstyle \sum^{\, \prime}}_{n=0}^{\lfloor N/2\rfloor-1}
               \left\{ 1 - p \pi \rho 
                       \sum\limits_{m=1}^\infty \left({\pi\over N}\right)^{2m}
                       {\Lambda_{2m+1}\over (2m+1)! } (n+\alpha)^{2m+1} 
                       \right\} \nonumber \\
   & & \qquad \qquad \qquad \qquad \times  
               \exp \{-2p[\pi \tau_0 \rho (n+\alpha) + i \pi \beta]\}
               \nonumber \\[2mm]
    & & \quad \quad -\real \sum\limits_{p=0}^\infty {1\over p}
           \mathop{\textstyle \sum^{\, \prime}}_{n=0}^{N-\lfloor N/2\rfloor-1}
                \left\{ 1 - p \pi \rho 
                       \sum\limits_{m=1}^\infty \left({\pi\over N}\right)^{2m}
                       {\Lambda_{2m+1}\over (2m+1)! } (n+1-\alpha)^{2m+1} 
                       \right\} \nonumber \\ 
   & & \qquad \qquad \qquad \qquad \qquad \qquad \times 
               \exp\{-2p[\pi \tau_0 \rho (n+1+\alpha) - i \pi \beta]\}
\label{def_f2_5}
\end{eqnarray}
where the $\Lambda_k$ are certain differential operators. The first ones are
\begin{subeqnarray}
\Lambda_3 &=& \Lambda_7 = 0 \\
\Lambda_5 &=& \lambda_5 \\
\Lambda_9 &=& \lambda_9  + {63\over 5} \lambda_5^2 
   {\partial \over \partial \lambda} \\
\Lambda_{11} &=& \lambda_{11} 
\label{def_Lambdas}
\end{subeqnarray} 

We can now extend the sum over $n$ to $n=\infty$ as the error is 
exponentially small. On the other hand, the contribution of the term with
$n+\alpha=N/2$ is also exponentially small, so we can take out this constraint.
Then, after rearranging the sums, we obtain 
\begin{eqnarray}
f_2 &=& \sum\limits_{n=0}^\infty \log\left| 
         1 - e^{-2\pi[\rho\tau_0 (n+\alpha)+i\beta]} \right| 
     + \sum\limits_{n=0}^\infty \log\left| 
         1 - e^{-2\pi[\rho\tau_0 (n+1-\alpha)-i\beta]} \right| \nonumber \\ 
    & & + \pi \rho \sum\limits_{m=1}^\infty 
        \left({\pi\over N}\right)^{2m} 
        {\Lambda_{2m+1}\over (2m+1)!} \real  
        \sum\limits_{p=1}^\infty \sum\limits_{n=0}^\infty 
        \left\{ (n+\alpha)^{2m+1} 
        e^{-2p\pi [\rho\tau_0 (n+\alpha)+i\beta] } \right. \nonumber \\ 
    & & \qquad \qquad \left.  
        + (n+1-\alpha)^{2m+1}    
        e^{-2p\pi[\rho\tau_0 (n+1-\alpha)+i\beta]} \right\} 
\label{def_f2_6}
\end{eqnarray}
The desired result can be obtained by plugging in 
\reff{formula1}/\reff{formulaK1}: 
\begin{eqnarray}
f_2 &=& \log\left| {\theta_{\alpha,\beta}(i\tau_0\rho)
                 \over\eta(i\tau_0\rho)} \right|  + 
      {\pi \lambda \rho \over 2} B_2(\alpha) \nonumber \\ 
     & & \qquad + 
      \pi \rho \sum\limits_{m=1}^\infty
      \left( {\pi \over N} \right)^{2m} {\Lambda_{2m+1}\over (2m+2)!}
      [B_{2m+2}(\alpha) - \real K_{2m+2}^{\alpha,\beta}(i\tau_0\rho)] 
\label{result_f2}
\end{eqnarray} 
where the elliptic $\theta$-function $\theta_{\alpha,\beta}$ and the 
Dedekind's $\eta$-function are defined in 
Appendix~\ref{sec_theta}, the objects $B_p(\alpha)$ are Bernoulli 
polynomials defined in Appendix~\ref{sec_Euler}, 
and $K_{2m+2}^{\alpha,\beta}$
are Kronecker's double series defined in Appendix~\ref{sec_Kronecker}.  
Then, the value of $Z_{\alpha,\beta}(0)$ is given by 
\begin{eqnarray}
\log Z_{\alpha,\beta}(0) &=& {N M\over 2} \log 2 + {NM \over 2\pi} 
      \int_0^\pi \omega_1(t) \, dt + \log\left| 
      {\theta_{\alpha,\beta}(i\tau_0\rho)
                   \over\eta(i\tau_0\rho)} \right| \nonumber \\
    & & \qquad - \pi \rho 
      \sum\limits_{m=1}^\infty
      \left( {\pi \over N} \right)^{2m} {\Lambda_{2m+1}\over (2m+2)!}
      \real K_{2m+2}^{\alpha,\beta}(i\tau_0\rho)
\label{result_Zab0}
\end{eqnarray}  

The free energy at the critical point can be computed directly 
from \reff{partition_function}:
\be
\label{free_energy}
f_c(N,M) = -{1\over V} \log 2 + {1\over 2}\log(2\sinh 2\beta_c) + 
            {1\over V}\log\sum_{\alpha,\beta} Z_{\alpha,\beta}(0)
\ee
The result \reff{result_Zab0} means that the free energy for both lattices
can be written as
\be
\label{series_free_energy}
f_c(N,\rho) = f_{\rm bulk} + \sum_{m=1}^\infty {f_{2m}(\rho)\over N^{2m}}
\ee
Thus, only even powers of $N^{-1}$ can occur, and in contrast to what 
happens in the square-lattice, we find some even powers whose coefficient 
vanishes (e.g., $f_4 = f_8 = 0$). The above result agrees with the 
formula found by Izmailian and Hu \cite{Izmailian3} for an Ising model 
on a $N\times\infty$ hexagonal (or triangular) lattice with periodic 
boundary conditions.  

The first coefficients for the triangular lattice are given by
\begin{subeqnarray}
\slabel{fbulk_tri}
f_{\rm bulk}^{\rm tri} 
                    &=& {1\over 2}\log {4\over \sqrt{3}} + 
                       {1\over 2\pi}\int_0^\pi \omega_1(t) \, dt 
                       \approx 0.8795853861\ldots \\[2mm]
f_2^{\rm tri}(\rho) &=& {1\over \rho} \log { 
          |\theta_2| + |\theta_3| + |\theta_4| \over 2 |\eta| } 
\slabel{f2_tri} \\
f_4^{\rm tri}(\rho) &=& f_8^{\rm tri}(\rho) = 0 \\
f_6^{\rm tri}(\rho) &=& - {\pi^5 \over 45 \sqrt{3}} \real 
  { |\theta_4| K^{{1\over2},0}_6 + |\theta_2| K^{0,{1\over2}}_6 
  + |\theta_3| K^{{1\over2},{1\over2}}_6 \over 
    |\theta_2| + |\theta_3| + |\theta_4| }
\slabel{f6_tri}
\end{subeqnarray}
where the $\theta_i$ are the standard $\theta$-functions defined in 
\reff{def_standard_thetas} and the functions $K^{\alpha,\beta}_6$ are
given in terms of $\theta$-functions in \reff{values_Kab_6}. As explained
in the Appendix~\ref{sec_theta} all the functions $\theta_i$, $\eta$ and
$K^{\alpha,\beta}_p$ are evaluated at $z=0$ and $\tau=i\tau_0 \rho$ 
\reff{shorthands_theta}. The numerical values of these coefficients for
several values of $\rho$ can be found in Table~\ref{table_free}.  

The coefficients of the hexagonal-lattice expansion are found to be 
\begin{subeqnarray}
\slabel{fbulk_hc}
f_{\rm bulk}^{\rm hc}   &=& {1\over 2}\log 2 \sqrt{6} +
                       {1\over 4\pi}\int_0^\pi \omega_1(t) \, dt
                       \approx 1.0250590964\ldots \\
f_{2m}^{\rm hc}(\rho) &=& {1\over 2} f_{2m}^{\rm tri}(\rho) 
\slabel{fm_hc}
\end{subeqnarray}
The numerical values of the coefficients $f_2^{\rm hc}$ and $f_6^{\rm hc}$ can
be obtained from Table~\ref{table_free} with the help of \reff{fm_hc}. 

\bigskip

\noindent
{\bf Remarks}. 
1. The values of the bulk critical free energy 
\reff{fbulk_tri}/\reff{fbulk_hc} indeed coincide with the values obtained 
from the well-known results in the thermodynamic limit 
\cite{Houtappel,Newell,Montroll} when $\beta=\beta_c$: 
\begin{subeqnarray}
f_{\rm bulk}^{\rm tri}(\beta) &=& 
  {1\over 2} \int_0^\pi \! \! \int_0^\pi {dx dy \over 4\pi^2} \log 
  [ \cosh^3 2\beta + \sinh^3 2\beta - 
          \omega(x,y) \sinh 2\beta ] \nonumber \\ 
  & & \qquad + \log 2 
  \slabel{f_bulk_tri_infty}
 \\[2mm]
f_{\rm bulk}^{\rm hc}(\beta) &=&  
  {1\over 4} \int_0^\pi \! \! \int_0^\pi {dx dy \over 4\pi^2} \log 
  [ 1 + \cosh^3 2\beta - \omega(x,y) \sinh^2 2\beta ]  \nonumber \\ 
   & & \qquad + {3\over 4} \log 2  
  \slabel{f_bulk_hc_infty}
\end{subeqnarray}
where 
\be
\label{def_omega}
\omega(x,y) = \cos x + \cos y + \cos(x-y)  
\ee

2. The limiting values of the coefficients $f_2$ and $f_6$ as $\rho\to\infty$
are easily found to be [c.f., \reff{theta_limits}]  
\begin{subeqnarray}
\lim_{\rho\to\infty} f_{2}^{\rm tri}(\rho) &=& {\sqrt{3}\pi \over 24} \\ 
\lim_{\rho\to\infty} f_{6}^{\rm tri}(\rho) &=& {31 \pi^5 \over 60480\sqrt{3}} 
\end{subeqnarray}
The corresponding limiting values for the hexagonal lattice are one half of 
the above values [c.f., \reff{fm_hc}].  

3. Using the properties of the $\theta$-functions 
\reff{def_Jacobi_transformation2}/\reff{identity_abs_theta} and 
of the functions $K_6^{\alpha,\beta}$ 
\reff{Jacobi_K6ab}/\reff{identity_re_K6ab} we can easily check that 
the terms \reff{f2_tri}/\reff{f6_tri} 
have the correct behavior under the transformation $N\leftrightarrow M$  
($\rho \to 1/\rho$). In particular,   
\begin{eqnarray}
f_2(\rho) &=& {f_2(1/\rho) \over \rho^2 } \\
f_6(\rho) &=& {f_6(1/\rho) \over \rho^6 } 
\end{eqnarray}

4. From \reff{result_Zab0}/\reff{def_Lambdas} we see that there is in 
general a non-zero contribution to $\log Z_{\alpha,\beta}(0)$ at 
any order $N^{-2m}$ with $m\ge 4$. However, we cannot rule out 
cancellations leading to the vanishing of any of the coefficients 
$f_{2m}(\rho)$ with $m\geq 5$ in \reff{series_free_energy}. Similar 
arguments apply to the other large-$N$ expansions in the next sections.

\begin{table}[htb]
\centering
\begin{tabular}{rll}
\hline\hline
\multicolumn{1}{c}{$\rho$}  & \multicolumn{1}{c}{$f_2^{\rm tri}(\rho)$} & 
                              \multicolumn{1}{c}{$f_6^{\rm tri}(\rho)$} \\
\hline
1  & 0.636514168294813 & 0.084178614254145 \\
2  & 0.340929552077890 & 0.052778553027830 \\
3  & 0.267452513800776 & 0.069393489802385 \\
4  & 0.242663080213048 & 0.079193473629707 \\
5  & 0.233284972993438 & 0.084621760086675 \\
6  & 0.229516370606439 & 0.087503667615417 \\ 
7  & 0.227941884733430 & 0.088999315024483 \\
8  & 0.227265427814348 & 0.089766374485276 \\
9  & 0.226968542233183 & 0.090157377673790 \\
10 & 0.226836041806366 & 0.090356069258841 \\
11 & 0.226776103731001 & 0.090456876341452 \\
12 & 0.226748689100100 & 0.090507980191895 \\
13 & 0.226736034656892 & 0.090533876582898 \\
14 & 0.226730148221756 & 0.090533876582898 \\
15 & 0.226727392017273 & 0.090553643011573 \\
16 & 0.226726094176590 & 0.090557009779414 \\
17 & 0.226725480047888 & 0.090558715190155 \\
18 & 0.226725188196890 & 0.090559579041296 \\
19 & 0.226725048974839 & 0.090560016609581 \\
20 & 0.226724982337581 & 0.090560238251165 \\  
$\infty$
   & 0.226724920529277 & 0.090560465757793 \\
\hline\hline
\end{tabular}
\caption{\protect\label{table_free}
Values of the coefficients $f_2^{\rm tri}(\rho)$ and $f_6^{\rm tri}(\rho)$  
for several values of the torus aspect ratio $\rho$.
}
\end{table}
     
%
%
\section{Finite-size-scaling corrections to the internal energy}
\label{sec_E}

Now we will deal with the internal energy \reff{def_energy}. Using 
\reff{partition_function}/\reff{properties_Zab} we can write the critical 
internal energy as follows:
\be
-E_c(N,\rho) = \coth 2\beta_c + {1\over V} 
  \left. {d\mu \over d\beta} \right|_{\beta=\beta_c} 
  {Z_{0,0}^\prime(0) \over \sum\limits_{\alpha,\beta} Z_{\alpha,\beta}(0)} 
\label{critical_energy}
\ee
The derivative $d\mu/d\beta$ can be easily computed from eq.~\reff{def_mass}. 
Thus, the only unknown object is $Z_{0,0}^\prime(0)$, which can be written as
\begin{eqnarray}
Z_{0,0}^\prime(0) &=& 2M \, 2^{NM/2} \,  
             \prod\limits_{n=0}^{N-1} \left( 1 + \sin^2 \phi_n + 
             \sin \phi_n \sqrt{3 + \sin^2\phi_n} \right)^{M/2} 
             \nonumber \\
  &  & \quad \times \prod\limits_{n=1}^{N-1} 
             \left| 1 - z_{-}(n,N,0)^M e^{M i \phi_n} \right| 
\label{def_Z00p} 
\end{eqnarray}
By noting that the first product is nothing more than $f_1$ \reff{def_f1} with  
$\alpha=0$, we can write \reff{def_Z00p} as
\begin{eqnarray}
\log Z_{0,0}^\prime(0) &=& {NM\over 2} \log 2 + \log 2M + 
             {NM \over 2\pi} \int_0^\pi \omega_1(t)\, dt  \nonumber \\
                       & & \qquad 
            - {\pi \rho\lambda\over 2} B_2(0) 
            - \pi\rho \sum\limits_{m=1}^\infty \left({\pi\over N}\right)^{2m}
              {B_{2m+2}(0)\over (2m+2)!} \lambda_{2m+1} \nonumber \\ 
  &  & \qquad + \sum\limits_{n=1}^{N-1}
             \log \left| 1 - z_{-}(n,N,0)^M e^{M i \phi_n} \right|
\label{def_Z00p_bis} 
\end{eqnarray}

The last sum in \reff{def_Z00p_bis} is equal to the definition of $f_2$ 
\reff{def_f2} with $\alpha=0$, except for the fact that the sum in 
\reff{def_Z00p_bis} starts at $n=1$ rather than at $n=0$. We can follow 
step by step the same procedure leading to \reff{def_f2_6}: the result 
coincides with \reff{def_f2_6} when $\alpha=0$ except that the first sum 
in \reff{def_f2_6} now starts at $n=1$. Using 
\reff{formula2}/\reff{formulaK1} we obtain the final result
\begin{eqnarray}
\log Z_{0,0}^\prime(0) &=& {NM\over 2}\log 2 + \log 2M + 
                           {NM \over 2\pi} \int_0^\pi \omega_1(t)\, dt
                           +2 \log|\eta(i\tau_0\rho)| \nonumber \\
                       & & \qquad
                - \pi \rho \sum\limits_{m=1}^\infty
           \left({\pi\over N}\right)^{2m} {\Lambda_{2m+1} \over (2m+2)!} 
           \real K_{2m+2}^{0,0}(i\tau_0\rho) 
\label{result_Z00P}
\end{eqnarray}
This equation implies that the critical internal energy can be 
written as a power series in $N^{-1}$:  
\be
\label{energy}
-E_c(N,\rho) = E_0 + \sum\limits_{m=0}^\infty {E_{2m+1}(\rho) \over N^{2m+1}}
\ee
For the triangular lattice we find that
\begin{subeqnarray}
E_0^{\rm tri} &=& 2 \\
\slabel{E1_tri}
E_1^{\rm tri}(\rho)
              &=& {3 |\theta_2 \theta_3 \theta_4 | \over 
                     |\theta_2| + |\theta_3| + |\theta_4| } \\
E_3^{\rm tri}(\rho)  &=& E_7^{\rm tri}(\rho)  = 0 \\
\slabel{E5_tri}
E_5^{\rm tri}(\rho)
              &=& -{\pi^5 \rho\over 15\sqrt{3}} 
                  {|\theta_2 \theta_3 \theta_4 | \over 
                   (|\theta_2| + |\theta_3| + |\theta_4|)^2}
              \real \left\{ (|\theta_2| + |\theta_3| + |\theta_4|) K_6^{0,0} 
                   \phantom{K^{{1\over2},{1\over2}}_6} 
                    \right. 
               \nonumber \\
              & & \qquad \qquad - \left.  
                |\theta_4| K^{{1\over2},0}_6 - |\theta_2| K^{0,{1\over2}}_6
                - |\theta_3| K^{{1\over2},{1\over2}}_6 \right\}  
\end{subeqnarray}
where we have used \reff{formula3}/\reff{values_Kab_6}. The numerical 
values of these coefficients can be found in Table~\ref{table_energy}.  
In the hexagonal-lattice case we obtain  
\begin{subeqnarray}
E_0^{\rm hc}            &=& {2\over \sqrt{3}} \\
E_{2m+1}^{\rm hc}(\rho) &=& {E_{2m+1}^{\rm tri}(\rho) \over 2\sqrt{3}}  
\slabel{Em_hc}
\end{subeqnarray}
The numerical values of the coefficients $E_1^{\rm hc}$ and $E_5^{\rm hc}$ can
be obtained from Table~\ref{table_energy} by using \reff{Em_hc}. 

\bigskip

\noindent
{\bf Remarks.}
1. The limiting values of the coefficients $E_1$ and $E_5$ as $\rho\to\infty$
are easily found to be [c.f., \reff{theta_limits}]
\be
\lim_{\rho\to\infty} E_{1}(\rho) =  
\lim_{\rho\to\infty} E_{5}(\rho) = 0  
\ee 
This formula is valid for the triangular and hexagonal lattices.
In particular, we expect that {\em all} the coefficients $E_{2m+1}(\rho)$ 
will vanish in the limit $\rho\to\infty$ due to the existence of the 
factor $|\theta_2\theta_3\theta_4|$ which vanishes exponentially fast. 
Thus, on an infinitely long torus, the internal energy for {\em any}
finite width $N$ is equal to the bulk value $E_0$ with no finite-size
corrections.

2. Using the properties of the $\theta$-functions
\reff{def_Jacobi_transformation2}/\reff{identity_abs_theta} and
of the functions $K_6^{\alpha,\beta}$
\reff{Jacobi_K6ab}/\reff{identity_re_K6ab} we can easily check that
the coefficients $E_1$ and $E_5$ \reff{E1_tri}/\reff{E5_tri}/\reff{Em_hc} 
have the correct behavior under the transformation $\rho \to 1/\rho$. 
In particular, 
\begin{eqnarray}
E_1(\rho) &=& {E_1(1/\rho) \over \rho   } \\
E_5(\rho) &=& {E_5(1/\rho) \over \rho^5 }
\end{eqnarray}

\begin{table}[htb]
\centering
\begin{tabular}{rll}
\hline\hline
\multicolumn{1}{c}{$\rho$}  & \multicolumn{1}{c}{$E_1^{\rm tri}(\rho)$} & 
                              \multicolumn{1}{c}{$E_5^{\rm tri}(\rho)$} \\
\hline
1  & 1.017408797595956 & -0.359705063388737 \\
2  & 0.612513647162813 & -0.178088378079924 \\
3  & 0.345040108164264 & -0.168599461543254 \\
4  & 0.185288835745847 & -0.127979167922216 \\
5  & 0.096804501605795 & -0.086206117890971 \\
6  & 0.049827662298672 & -0.054108487929080 \\ 
7  & 0.025447703091251 & -0.032506071497113 \\
8  & 0.012944169002509 & -0.018975959727317 \\
9  & 0.006570580061525 & -0.010859541565321 \\
10 & 0.003331786807789 & -0.006125084912961 \\
11 & 0.001688570266906 & -0.003416526641140 \\
12 & 0.000855546533105 & -0.001888941526185 \\
13 & 0.000433419665204 & -0.001036828005370 \\
14 & 0.000219555049642 & -0.000565662233279 \\
15 & 0.000111214898315 & -0.000307012198010 \\
16 & 0.000056334542069 & -0.000165883847105 \\
17 & 0.000028535313425 & -0.000089278100310 \\
18 & 0.000014454016292 & -0.000047882460122 \\
19 & 0.000007321388062 & -0.000025601385603 \\
20 & 0.000003708495908 & -0.000013650380771 \\  
$\infty$
   & 0  & \phantom{-}0  \\
\hline\hline
\end{tabular}
\caption{\protect\label{table_energy}
Values of the coefficients $E_1^{\rm tri}(\rho)$ and $E_5^{\rm tri}(\rho)$  
for several values of the torus aspect ratio $\rho$.
}
\end{table}
 
%
%
\section{Finite-size-scaling corrections to the specific heat}
\label{sec_CH}

The specific heat at criticality is given by the following formula 
\begin{eqnarray}
C_{H,c} &=& {-2\over\sinh^2 2\beta_c} + {1\over V} 
            \left. {d^2 \mu\over d\beta^2}\right|_{\beta=\beta_c} 
            {Z_{0,0}^\prime(0) \over \sum\limits_{\alpha,\beta} 
             Z_{\alpha,\beta}(0)} \nonumber \\
        & & \qquad +  
            {1\over V}
            \left. {d \mu\over d\beta}\right|_{\beta=\beta_c}^2
            \left[ 
            {\sum\limits_{\alpha,\beta} Z_{\alpha,\beta}^{\prime\prime}(0) 
             \over 
             \sum\limits_{\alpha,\beta} Z_{\alpha,\beta}(0)} - 
             \left( 
             {Z_{0,0}^\prime(0) \over \sum\limits_{\alpha,\beta}
             Z_{\alpha,\beta}(0)} \right)^2 \, \right]
\label{def_CH}
\end{eqnarray}
The main goal of this section is to compute the ratio 
\be
{\sum\limits_{\alpha,\beta} Z_{\alpha,\beta}^{\prime\prime}(0)
             \over
             \sum\limits_{\alpha,\beta} Z_{\alpha,\beta}(0)}
\label{ratio_zpp_over_z}
\ee
where the sums go over $(\alpha,\beta)\neq (0,0)$. After some algebra, we 
can write the derivative $Z_{\alpha,\beta}^{\prime\prime}(0)$ as follows
\be
Z_{\alpha,\beta}^{\prime\prime}(0) = {4 M N \over \pi\sqrt{3}} 
    Z_{\alpha,\beta}(0)  \left[ {\cal S}^{(1)}_\alpha + 
                               2{\cal S}^{(2)}_{\alpha,\beta} 
    + {\pi\sqrt{3}\over 4} \rho \, \delta_{\alpha,0} \right] 
\label{def_ZPP0} 
\ee
where the sums ${\cal S}^{(j)}$ are given by 
\begin{subeqnarray}
{\cal S}^{(1)}_\alpha &=& {\pi\sqrt{3}\over 2N} 
                          \sum\limits_{n=\delta_{\alpha,0}}^{N-1} 
{1\over \sin \phi_{n+\alpha} \sqrt{ 3 + \sin^2 \phi_{n+\alpha} }} 
\slabel{def_s1} \\
{\cal S}^{(2)}_{\alpha,\beta} &=& {\pi\sqrt{3}\over 2N} \real 
                          \sum\limits_{n=\delta_{\alpha,0}}^{N-1} 
{1\over \sin \phi_{n+\alpha} \sqrt{ 3 + \sin^2 \phi_{n+\alpha} }} 
{z_{-}^M e^{-2\pi i \beta + M i \phi_{n+\alpha} } \over 
1 - z_{-}^M e^{-2\pi i \beta + M i \phi_{n+\alpha} } } 
\slabel{def_s2}
\label{def_s}
\end{subeqnarray}
The variables $\phi_{n+\alpha}$ and $z_{-} = z_{-}(n+\alpha,N,0)$ are given by 
\reff{def_zpm_bis} and $\delta_{\alpha,0}$ is the usual Kronecker's delta.  

The first step is to compute the sum ${\cal S}^{(1)}_\alpha$ 
\reff{def_s1}. We will follow a procedure similar to the one used in 
ref.~\cite{Salas_01} for the square lattice. Let us define the function 
\be
\omega_3(k) = { \sqrt{3} \over \sin k \sqrt{3 + \sin^2 k}} - {1\over k} 
              + {1\over k-\pi}
\label{def_w3}
\ee
This function and all its derivatives are integrable over the interval 
$[0,\pi]$ so we can apply the Euler-MacLaurin formula 
\reff{Euler_MacLaurin_formula_final}. The final result is 
\be
{\cal S}^{(1)}_\alpha(N) = \sum\limits_{n=\delta_{\alpha,0}}^{N-1} 
{1 \over n+\alpha} + {1\over 2N} \delta_{\alpha,0} + 
{1\over 2} \int_0^\pi \omega_3(t) \, dt - 
\sum\limits_{m=1}^\infty \left({\pi\over N}\right)^{2m} 
{B_{2m}(\alpha)\over (2m)!} \widetilde{\gamma}_{2m-1}
\label{def_s1_2}
\ee
where the coefficients $\widetilde{\gamma}_{2m-1}$ come from the expansion of
$\omega_3(k)$ in powers of $k$:
\begin{subeqnarray}
\omega_3(k) &=& \sum\limits_{m=0}^\infty { \widetilde{\gamma}_{m} \over m!} \,
                k^m 
\slabel{def_gammatilde} \\
            &=& - \sum\limits_{m=0}^\infty {k^{m} \over \pi^{m+1} } 
                + \sum\limits_{m=1}^\infty {\gamma_{2m+1} \over (2m+1)! } \, 
                  k^{2m+1} 
\slabel{def_gamma} 
\end{subeqnarray}
In general, the coefficient $\widetilde{\gamma}_m$ contains two 
contributions: one comes from the term $1/(k-\pi)$ which gives the (trivial)
coefficient $-\pi^{-(m+1)} m!$, and the other contribution comes from the 
first two terms in the l.h.s.\  of \reff{def_w3}. We will denote by $\gamma_m$ 
this latter (non-trivial) contribution. In particular, 
only the coefficients $\gamma_{2m+1}$ with $m=1,2,3,\ldots$ are non-zero. 
The first non-vanishing coefficients $\gamma_m$ are
\be
\gamma_3 =  {8  \over 15} \,; \qquad 
\gamma_5 = -{80 \over 21} \,; \qquad 
\gamma_7 =  {448\over 5} 
\label{values_gammas}
\ee
On the other hand, the value of the integral in \reff{def_s1_2} is 
\be
{1\over 2} \int_0^\pi \omega_3(t) \, dt = \log {\sqrt{3}\over \pi}  
\label{integral_w3}
\ee

In computing the sums $\sum_{n=\delta_{\alpha,0}}^{N-1} (n+\alpha)^{-1}$ 
we will use the result (See e.g., \cite{Caracciolo_98})
\be
\sum\limits_{n=1}^{N} {1\over N} = \log N + \gamma_E + {1\over 2N} 
  - \sum\limits_{k=1}^\infty {B_{2k} \over 2k} {1\over N^{2k}} 
\label{series_one_over_N}
\ee
(where $\gamma_E \approx 0.5772156649$ is the Euler constant) 
and take into account that $\alpha=0,1/2$. In the simplest case $\alpha=0$  
we have 
\begin{eqnarray}
{\cal S}^{(1)}_0(N) &=& \log N + \gamma_E + \log {\sqrt{3}\over \pi} -   
\sum\limits_{m=1}^\infty \left({\pi\over N}\right)^{2m}
{B_{2m}\over (2m)!} \widetilde{\gamma}_{2m-1} + {1\over 2N} \nonumber \\
                    & & \qquad 
+ {1\over 2(N-1)} + \log\left( 1 - {1\over N}\right) - 
\sum\limits_{m=1}^\infty {B_{2m} \over 2m} {1 \over (N-1)^{2m} }
\label{def_s10_3}
\end{eqnarray}
This expression can be simplified by expanding it in powers of $N^{-1}$, 
and then using formulas \reff{relation1}/\reff{relation2}. A further 
simplification can be made if we take into account \reff{def_gamma}.
The final result for $\alpha=0$ is 
\be
{\cal S}^{(1)}_0(N) = \log N + \gamma_E + \log {\sqrt{3}\over \pi} -
\sum\limits_{m=2}^\infty \left({\pi\over N}\right)^{2m}
{B_{2m}\over (2m)!} \gamma_{2m-1} 
\ee
The value for $\alpha=1/2$ can be obtained using similar arguments 
in addition to \reff{B_onehalf}. The final result for ${\cal S}^{(1)}_\alpha$ 
is
\be
{\cal S}^{(1)}_\alpha(N) = \log N + \gamma_E + \log {4\sqrt{3}\over \pi} -
\log 4 \, \delta_{\alpha,0} - 
\sum\limits_{m=2}^\infty \left({\pi\over N}\right)^{2m}
{B_{2m}(\alpha)\over (2m)!} \gamma_{2m-1}
\label{s1_final}
\ee
In the above result only enter the non-trivial Taylor coefficients of the
function $\omega_3$.

The second step is to compute the sums ${\cal S}^{(2)}_{\alpha,\beta}$ 
\reff{def_s2}. The procedure is similar to the ones already done in 
Sections~\ref{sec_F} and \ref{sec_E}. We first write 
\be
z_{-}(n+\alpha,N,0)^M = e^{-M \log z_{-}(n+\alpha,N,0)} = 
                        e^{-M \omega_2(\phi_{n+\alpha})}
\ee
where the function $\omega_2$ has been defined in \reff{def_w2}. Then we
split the sum $\sum_{n=\delta_{\alpha,0}}^{N-1}$ into two parts: 
$n\in [\delta_{\alpha,0},\lfloor N/2 \rfloor -1]$ and 
$n\in [\lfloor N/2 \rfloor,N-1]$. 
In the second sum we perform the 
change $n\to N-1-n$ and using the properties of $\omega_3$ we arrive at 
\begin{eqnarray}
{\cal S}^{(2)}_{\alpha,\beta} &=& {\pi \sqrt{3}\over 2N} \real \left[  
\sum\limits_{n=\delta_{\alpha,0}}^{\lfloor N/2 \rfloor -1}
{1\over \sin \phi_{n+\alpha} \sqrt{ 3 + \sin^2 \phi_{n+\alpha} }}
{e^{-2[M(\omega_2(\phi_{n+\alpha})
                      - i \phi_{n+\alpha})/2 + \pi i \beta] } \over
1 - e^{-2[M(\omega_2(\phi_{n+\alpha})
                      - i \phi_{n+\alpha})/2 + \pi i \beta] } } \right. 
  \nonumber \\
& & \left. \qquad \qquad + 
\sum\limits_{n=0}^{N-\lfloor N/2 \rfloor -1}
\left( \begin{array}{l}
     \alpha\to 1-\alpha \\
     \beta \to -\beta 
       \end{array} \right) \right] 
\label{def_s2_2}
\end{eqnarray}  
where the second term is the same as the first one with 
$(\alpha,\beta)$ replaced by ($1-\alpha,-\beta)$. Now we perform several
Taylor expansions: first, we expand the denominator $1 - e^{-2A}$ in powers of 
$e^{-2A}$:  
\begin{eqnarray}
{\cal S}^{(2)}_{\alpha,\beta} &=& {\pi \sqrt{3}\over 2N} \real \left[
\sum\limits_{n=\delta_{\alpha,0}}^{\lfloor N/2 \rfloor -1}
\sum\limits_{p=1}^\infty 
{e^{-2p[M(\omega_2(\phi_{n+\alpha})
                      - i \phi_{n+\alpha})/2 + \pi i \beta] }  \over 
   \sin \phi_{n+\alpha} \sqrt{ 3 + \sin^2 \phi_{n+\alpha} } } \right.
  \nonumber \\
& & \left. \qquad \qquad +
\sum\limits_{n=0}^{N-\lfloor N/2 \rfloor -1}
\sum\limits_{p=1}^\infty
{e^{-2p[M(\omega_2(\phi_{n+1-\alpha}) 
               - i \phi_{n+1-\alpha})/2 - \pi i \beta] }  \over
   \sin \phi_{n+1-\alpha} \sqrt{ 3 + \sin^2 \phi_{n+1-\alpha} } }
        \right]
\label{def_s2_3}
\end{eqnarray}
Secondly, we expand $e^{-2p (M\omega_2/2)}$ as we did in \reff{def_f2_4}
and finally, we expand the function 
\be
{\sqrt{3} \over \sin k \sqrt{3+\sin^2 k}} = 
  \omega_3(k) + {1\over k} - {1\over k -\pi} = 
   {1\over k} + \sum\limits_{m=1}^\infty
   {\gamma_{2m+1}\over (2m+1)!} k^{2m+1}  
\ee
in powers of $k$. After rearranging the series, extending the sums over $n$ 
to $\infty$ (as the error is exponentially small) and using  
\reff{relation_log_bis}/\reff{formulaK1} we obtain
\begin{eqnarray}
{\cal S}^{(2)}_{\alpha,\beta} &=& - \real \log \theta_{\alpha,\beta}  
  + \left[ \log 2 - {\pi \rho \sqrt{3}\over 8} \right] \delta_{\alpha,0} 
  \nonumber \\
 & & \qquad + {1\over 2}\sum\limits_{k=1}^\infty 
     \left({\pi \over N}\right)^{2k+2} {\gamma_{2k+1}\over (2k+2)!} 
     \left[B_{2k+2}(\alpha)- \real K_{2k+2}^{\alpha,\beta}(i\tau_0\rho)
     \right] \nonumber \\
 & & \qquad - {\pi\rho\over 2} \sum\limits_{k,m=1}^\infty
\left({\pi \over N}\right)^{2m+2k+2} {\Lambda_{2m+1}\over (2m+1)!}
     {\gamma_{2k+1}\over (2k+2)!} 
     W_{2m+2k+2}^{\alpha,\beta}(i\tau_0\rho) \nonumber \\
 & & \qquad - {\pi\rho\over 2} \sum\limits_{m=1}^\infty 
\left({\pi \over N}\right)^{2m} {\Lambda_{2m+1}\over (2m+1)!}
     W_{2m}^{\alpha,\beta}(i\tau_0\rho)
\label{s2_final}
\end{eqnarray}
where the function $W_{m}^{\alpha,\beta}(\tau)$ is defined as follows
\begin{eqnarray}
W_{m}^{\alpha,\beta}(\tau) &=& \real \sum_{n=0}^\infty \left[  
      (n+\alpha)^m {e^{2\pi i (\tau(n+\alpha)-\beta)} \over 
                   (1-e^{2\pi i (\tau(n+\alpha)-\beta)})^2 } \right. 
   \nonumber \\
 & & \qquad \left. + 
(n+1-\alpha)^m {e^{2\pi i (\tau(n+1-\alpha)+\beta)} \over 
                   (1-e^{2\pi i (\tau(n+1-\alpha)+\beta)})^2 } \right] 
\label{def_W}
\end{eqnarray}

The ratio $Z_{\alpha,\beta}^{\prime\prime}(0)/Z_{\alpha,\beta}(0)$ 
\reff{def_ZPP0} can be written as a power series in $N^{-1}$: 
\be
{1\over M N} {Z_{\alpha,\beta}^{\prime\prime}(0) \over Z_{\alpha,\beta}(0)}
= {4\over \pi\sqrt{3}} \left[ \log N + \gamma_E +
\log {4\sqrt{3}\over 2} - 2 \real \log \theta_{\alpha,\beta} \right] 
+ \sum\limits_{m=2}^\infty {\widetilde{d}_{2m}^{\alpha,\beta}(\rho) \over 
N^{2m}} 
\label{ratio_ZPP_final}
\ee
This series contains only {\em even} powers of $N^{-1}$ and it starts at 
$N^{-4}$ (i.e., $\widetilde{d}_2^{\alpha,\beta}=0$). The first non-vanishing 
coefficient $\widetilde{d}_{2m}^{\alpha,\beta}$ is 
\be
\widetilde{d}_4^{\alpha,\beta}(\rho) = 
  - { \pi^4    \over 45}   \real K_4^{\alpha,\beta}(i\tau_0\rho) 
  - {2\pi^5\rho\over 15\sqrt{3}} W_4^{\alpha,\beta}(i\tau_0\rho)
\label{def_d4tilde}
\ee
It is worth noticing that the terms with $\delta_{\alpha,0}$ in 
\reff{def_ZPP0}/\reff{s1_final}/\reff{s2_final} cancel out exactly. 

The computation of the ratio \reff{ratio_zpp_over_z} is straightforward 
from \reff{result_Zab0}/\reff{ratio_ZPP_final}. The leading term grows 
like $\log N$ and the rest can be expressed as a power series in $N^{-1}$ 
where only {\em even} powers of $N^{-1}$ enter:  
\be
{1\over MN} { \sum\limits_{\alpha,\beta} Z_{\alpha,\beta}^{\prime\prime}(0)
              \over
              \sum\limits_{\alpha,\beta} Z_{\alpha,\beta}(0) } =
{4\over \pi\sqrt{3}} \log N + d_0(\rho) + 
   \sum\limits_{m=2}^\infty {d_{2m}(\rho) \over N^{2m} } 
\label{ratio_zpp_over_z_final}
\ee
The coefficient associated to $N^{-2}$ vanishes, so the 
first two non-zero coefficients $d_m(\rho)$ are 
\begin{subeqnarray}
\slabel{d0_final}
d_0(\rho) &=& {4\over \pi\sqrt{3}} \left[ \gamma_E + 
               \log {4\sqrt{3}\over\pi} - 2  
   {\sum |\theta_i| \real \log \theta_i 
   \over \sum |\theta_i| } \right] \\
d_4(\rho) &=& -{4\pi^3\over 45} \left[ {2\pi\rho\over3} \left\{ 
{(\sum |\theta_i| \real \log \theta_i)
 (\sum |\theta_{\alpha,\beta}| \real K_6^{\alpha,\beta}) \over 
 (\sum |\theta_i|)^2} \right. \right. \nonumber \\ 
  & & \qquad \qquad \qquad - \left. \left.  
 {\sum |\theta_{\alpha,\beta}| \real K_6^{\alpha,\beta} 
                             \real \log \theta_{\alpha,\beta} \over 
  \sum |\theta_i|} \right\} \right. \nonumber \\
          & & \qquad + \left. {1\over \sqrt{3}} 
{\sum |\theta_{\alpha,\beta}| \real K_4^{\alpha,\beta}) \over 
 \sum |\theta_i|} + 
2\pi\rho\, 
{\sum |\theta_{\alpha,\beta}| W_4^{\alpha,\beta} \over \sum |\theta_i| }
\right] 
\slabel{d4_final}
\end{subeqnarray}
where we have denoted by $\theta_i$ the $\theta$-functions in the standard 
notation \reff{def_standard_thetas}. The numerical values of these 
coefficients can be found in Table~\ref{table_ratio_Z}. 

\bigskip

\noindent
{\bf Remarks.}
1. The limiting values of the coefficients $d_0$ and $d_4$ as $\rho\to\infty$
are easily found to be [c.f., \reff{theta_limits}]
\begin{subeqnarray}
\slabel{lim_d0}
\lim_{\rho\to\infty} d_{0}(\rho) &=& {4\over\pi\sqrt{3}}\left[\gamma_E + 
         \log{4\sqrt{3}\over \pi} \right] \\
\lim_{\rho\to\infty} d_{4}(\rho) &=& - {7\pi^3 \over 2700\sqrt{3}}  
\end{subeqnarray}

2. Using the properties of the $\theta$-functions
\reff{def_Jacobi_transformation2}/\reff{identity_abs_theta} 
we can easily verify that $d_0(\rho)$ has the right behavior under 
the transformation $N\leftrightarrow M$ ($\rho \to 1/\rho$): 
\be
{4\over\pi\sqrt{3}}\log \rho + d_0(1/\rho) = d_0(\rho) 
\ee
The behavior of $d_4(\rho)$ under this transformation can be checked 
numerically to be the right one  
\be
d_4(\rho) = {d_4(1/\rho) \over \rho^4}  
\ee

\bigskip

\begin{table}[htb]
\centering
\begin{tabular}{rll}
\hline\hline
\multicolumn{1}{c}{$\rho$}  & \multicolumn{1}{c}{$d_0(\rho)$} &
                              \multicolumn{1}{c}{$d_4(\rho)$} \\
\hline
1  &0.993000152525293 &-0.034652876469773  \\
2  &1.205930021583709 &-0.084727027938228  \\
3  &1.233520243783654 &-0.146295429270869  \\
4  &1.189798214112785 &-0.167434330275211  \\
5  &1.134144577781982 &-0.157595538508037  \\
6  &1.088416663135744 &-0.134596242800881  \\
7  &1.056420958518946 &-0.110373125092552  \\
8  &1.035808103247928 &-0.090138741047513  \\
9  &1.023167108495542 &-0.075083168632513  \\
10 &1.015661512376353 &-0.064637113563536  \\
11 &1.011305166086030 &-0.057721974432342  \\
12 &1.008818898345350 &-0.053296998167206  \\
13 &1.007418256678703 &-0.050537602364217  \\
14 &1.006637342854852 &-0.048851578543533  \\
15 &1.006205629259680 &-0.047838335036578  \\
16 &1.005968648007326 &-0.047237747222882  \\
17 &1.005839340365928 &-0.046885886437925  \\
18 &1.005769147640350 &-0.046681800486309  \\
19 &1.005731215221337 &-0.046564452961825  \\
20 &1.005710797002295 &-0.046497492807615  \\
$\infty$
   &1.005687333437919 & -0.046411250116879  \\
\hline\hline
\end{tabular}
\caption{\protect\label{table_ratio_Z}
Values of the coefficients $d_0(\rho)$ and $d_4(\rho)$
for several values of the torus aspect ratio $\rho$.
}
\end{table}

The specific heat for the triangular and hexagonal lattices can be 
obtained from \reff{def_CH} and using the results of this section and of
Section~\ref{sec_E}. In particular, we can write for both lattices
\be
C_{H,c}(N,\rho) = C_{00} \log N + C_0(\rho) + \sum\limits_{m=1}^\infty
  {C_m(\rho)\over N^m} 
\label{CH_final}
\ee

For the triangular lattice the first coefficients are given by
\begin{subeqnarray}
C_{00}^{\rm tri}\phantom{(\rho)} 
                    &=& {12\sqrt{3}\over \pi} \\
C_0^{\rm tri}(\rho) &=& 9 \, d_0(\rho) - 6 - \rho\,  E_1^{\rm tri}(\rho)^2 \\
C_1^{\rm tri}(\rho) &=& -2\,  E_1^{\rm tri}(\rho) \\
C_2^{\rm tri}(\rho) &=& C_3^{\rm tri}(\rho) = 0\\
C_4^{\rm tri}(\rho) &=& 9 \, d_4(\rho) - 
       2 \, \rho\,  E_1^{\rm tri}(\rho)\,  E_5^{\rm tri}(\rho) \\
C_5^{\rm tri}(\rho) &=& -2\,  E_5^{\rm tri}(\rho)
\label{C_tri_final}
\end{subeqnarray}
and for the hexagonal lattice the corresponding coefficients are
\begin{subeqnarray}
C_{00}^{\rm hc}\phantom{(\rho)}
                   &=& {2\sqrt{3}\over \pi} \\
C_0^{\rm hc}(\rho) &=& {3\over 2}\,  d_0(\rho) - {2\over 3}\,  
                       - 2\,  \rho\,  E_1^{\rm hc}(\rho)^2 \\
C_1^{\rm hc}(\rho) &=& -{2\over\sqrt{3}}\,  E_1^{\rm hc}(\rho) \\
C_2^{\rm hc}(\rho) &=& C_3^{\rm hc}(\rho) = 0\\
\slabel{C4_hc_final}
C_4^{\rm hc}(\rho) &=& {3\over 2}\,  d_4(\rho) - 
       4\,  \rho\,  E_1^{\rm hc}(\rho)\,  E_5^{\rm hc}(\rho) 
                    = {1\over 6} C_4^{\rm tri}(\rho) \\
C_5^{\rm hc}(\rho) &=& -{2\over\sqrt{3}}\,  E_5^{\rm hc}(\rho)
\label{C_hc_final}
\end{subeqnarray}
The numerical values of the coefficients $C_0^{\rm tri}$, $C_4^{\rm tri}$ and
$C_0^{\rm hc}$ can be found in Table~\ref{table_CH}. The values of the 
coefficients $C_1$ and $C_5$ can be obtained from Table~\ref{table_energy}, 
and the value of $C_4^{\rm hc}$ can be read from Table~\ref{table_CH} with
the help of \reff{C4_hc_final}. 

\begin{table}[htb]
\centering
\begin{tabular}{rlll}
\hline\hline
\multicolumn{1}{c}{$\rho$}  & \multicolumn{1}{c}{$C_0^{\rm tri}(\rho)$} &
                              \multicolumn{1}{c}{$C_4^{\rm tri}(\rho)$} & 
                              \multicolumn{1}{c}{$C_0^{\rm hc}(\rho)$}  \\
\hline
1  &1.901880711301990 & \phantom{-}0.420058303835062 &0.650313451883665 \\
2  &4.103024258331998 & -0.326217003543879           &1.017170709722000 \\ 
3  &4.744524165326869 & -0.967617404753895           &1.124087360887811 \\ 
4  &4.570856116406857 & -1.317204084284658           &1.095142686067810 \\ 
5  &4.160445642382107 & -1.334908443794273           &1.026740940397018 \\ 
6  &3.780853192640795 & -1.179012991639663           &0.963475532106799 \\ 
7  &3.503255527522170 & -0.981777257847270           &0.917209254587028 \\ 
8  &3.320932517142029 & -0.807318620952495           &0.886822086190338 \\ 
9  &3.208115423758777 & -0.674464154921458           &0.868019237293130 \\ 
10 &3.140842603353847 & -0.581325872529628           &0.856807100558975 \\ 
11 &3.101715130809265 & -0.519370850894424           &0.850285855134878 \\ 
12 &3.079361301589703 & -0.479634197647875           &0.846560216931617 \\ 
13 &3.066761868024451 & -0.454826737355134           &0.844460311337408 \\ 
14 &3.059735410831789 & -0.439660729459804           &0.843289235138631 \\ 
15 &3.055850477805811 & -0.430543990999289           &0.842641746300969 \\ 
16 &3.053717781288642 & -0.425139425966243           &0.842286296881440 \\ 
17 &3.052554049450865 & -0.421972891323654           &0.842092341575144 \\ 
18 &3.051922325002618 & -0.420136179461406           &0.841987054167103 \\ 
19 &3.051580935973584 & -0.419080069533795           &0.841930155995597 \\ 
20 &3.051397172745595 & -0.418477433243636           &0.841899528790933 \\ 
$\infty$                                            
   &3.051186000941275 & -0.417701251051913           &0.841864333490213 \\ 
\hline\hline
\end{tabular}
\caption{\protect\label{table_CH}
Values of the coefficients $C_0^{\rm tri}(\rho)$, $C_4^{\rm tri}(\rho)$ and
$C_0^{\rm hc}(\rho)$ for several values of the torus aspect ratio $\rho$.
}
\end{table}

\bigskip

\noindent
{\bf Remarks.}
1. The fact that the coefficients $C_1$ and $C_3$ are proportional respectively
to $E_1$ and $E_5$ for the triangular \reff{C_tri_final} and hexagonal 
\reff{C_hc_final} lattices is not accidental. In fact, from 
\reff{def_CH}/\reff{ratio_zpp_over_z_final} we conclude that {\em all} the
{\em odd} coefficients in the specific-heat expansion are proportional to
the corresponding coefficients of the internal-energy expansion. 
In fact, the proportional constant is given by 
[See \reff{def_CH}/\reff{critical_energy}]
\be
{E_{2m+1} \over C_{2m+1}} = \left. {d\mu \over d\beta}\right|_{\beta=\beta_c}  
                     \left( \left. {d^2\mu \over d\beta^2}
                            \right|_{\beta=\beta_c}
                      \right)^{-1} 
\label{proportional_law}
\ee
Indeed, for $m=1,3$ this ratio is indeterminate as both coefficients vanish.

2. The limiting values of the coefficients $C_m(\rho)$ as $\rho\to\infty$
are easily found to be [c.f., \reff{theta_limits}]
\begin{subeqnarray}
\lim_{\rho\to\infty} C_{0}^{\rm tri}(\rho) &=& 
{12\sqrt{3}\over\pi} \left[\gamma_E +
         \log{4\sqrt{3}\over \pi} - {\pi\over 2\sqrt{3}}\right] \\
\lim_{\rho\to\infty} C_{1}^{\rm tri}(\rho) &=&  
\lim_{\rho\to\infty} C_{5}^{\rm tri}(\rho) = 0 \\
\lim_{\rho\to\infty} C_{4}^{\rm tri}(\rho) &=& - {7\pi^3 \over 300\sqrt{3}}
\end{subeqnarray}
\begin{subeqnarray}
\lim_{\rho\to\infty} C_{0}^{\rm hc}(\rho) &=& 
{2\sqrt{3}\over\pi} \left[\gamma_E +
         \log{4\sqrt{3}\over \pi} - {\pi\over 3\sqrt{3}}\right]\\
\lim_{\rho\to\infty} C_{1}^{\rm hc}(\rho) &=&  
\lim_{\rho\to\infty} C_{5}^{\rm hc}(\rho) = 0 \\
\lim_{\rho\to\infty} C_{4}^{\rm hc}(\rho) &=& - {7\pi^3 \over 1800\sqrt{3}}
\end{subeqnarray}

3. The behavior of the coefficients $C_m(\rho)$ under the transformation
$\rho\to 1/\rho$ is the expected one  
\begin{subeqnarray}
C_0(\rho) &=& C_{00} \log \rho + C_0(1/\rho) \\ 
C_m(\rho) &=& {C_m(1/\rho) \over \rho^m } \qquad \qquad \quad \quad  
              \hbox{\rm for $m\geq 1$}  
\end{subeqnarray}

4. From Table~\ref{table_CH} it is clear that $C_4^{\rm tri}$ should 
vanish at a value between 1 and 2. Actually, due to (\ref{C_hc_final}e),
$C_4^{\rm hc}$ should also vanish at the same value of $\rho$.  
We have found numerically that $C_4$ vanishes at
\be
\rho_{\rm min} \approx 1.4688897779
\ee
Indeed, due to the transformation properties of $C_4(\rho)$ under
the transformation $\rho\to1/\rho$, $C_4$ also vanishes at 
$\rho_{\rm min}^{-1} \approx 0.6807862748$. This is similar to
what happens in the square lattice \cite{Salas_01}. 

\bigskip

%
%
\section{Higher derivatives of the free energy} 
\label{sec_der_CH}

\subsection{Finite-size-scaling corrections to $f^{(3)}_c$} \label{sec_f3}

In this section we will consider the third derivative of the free energy 
\reff{def_der_CH} at criticality. Even though this observable is not 
relevant in practice, its computation is interesting as it provides
new insights into the finite-size-scaling function $\widehat{W}$ defined in
Section~\ref{sec_irrel}. 
The observable $f^{(3)}_c$ \reff{def_der_CH} can be written as follows:
\begin{eqnarray}
f^{(3)}_c &=& 
{8\cosh 2\beta_c \over \sinh^3 \beta_c} + 
{1\over V} \left.{d^3 \mu \over d\beta^3}\right|_{\beta=\beta_c} 
{Z_{00}^\prime(0)\over \sum Z_{\alpha,\beta}(0)} 
+ {1\over V} \left( {d\mu \over d\beta}\right)_{\beta=\beta_c}^3 
\nonumber \\ 
  & & \qquad \times  
\left[ 
{Z_{\alpha,\beta}'''(0) \over \sum Z_{\alpha,\beta}(0)}
- 3 
{\sum Z_{\alpha,\beta}''(0) \over \sum Z_{\alpha,\beta}(0)} 
{Z_{00}'(0) \over \sum Z_{\alpha,\beta}(0)} + 
2 \left( {Z_{00}'(0) \over \sum Z_{\alpha,\beta}(0)} \right)^3 \right] 
\nonumber \\
  & & + 
{3\over V} \left.{d^2 \mu \over d\beta^2}\right|_{\beta=\beta_c}  
\left.{d\mu \over d\beta}\right|_{\beta=\beta_c}
\left[ {\sum Z_{\alpha,\beta}''(0) \over \sum Z_{\alpha,\beta}(0)} - 
\left( {Z_{00}'(0) \over \sum Z_{\alpha,\beta}(0)} \right)^2 
\right] 
\label{def_der_CH_bis}
\end{eqnarray}
The only unknown object is the derivative $Z_{0,0}'''(0)$, which can be 
written in the following way
\be
{Z_{0,0}'''(0) \over Z_{0,0}'(0)} = M^2 + {12 M N \over \pi\sqrt{3} }
\left[ {\cal S}_0^{(1)} + 2 {\cal S}_{0,0}^{(2)} \right] 
\label{def_ZPPP}
\ee
where the sums ${\cal S}^{(j)}$ were defined in 
\reff{def_s}. By following step by step the procedure 
developed in Section~\ref{sec_CH} and leading to \reff{ratio_ZPP_final}, we 
can compute the finite-size expansion of the ratio \reff{def_ZPPP} 
\be
{1\over M N} {Z_{0,0}'''(0) \over Z_{0,0}'(0)} = 
{12 \over \pi \sqrt{3}} \log N + \widetilde{A}(\rho) + 
\sum\limits_{m=2}^\infty {\widetilde{A}_{2m}\over N^{2m}} 
\label{final_ZPPP}
\ee

By plugging in \reff{def_der_CH_bis} the above result \reff{final_ZPPP} and  
the results already obtained in Sections~\ref{sec_E} and \ref{sec_CH}, 
we obtain  
\be
f^{(3)}_c(N,\rho) = {\cal A}_1(\rho) N + A_{00} \log N + A_0(\rho) + 
\sum\limits_{m=1}^\infty {A_m(\rho)\over N^m}
\label{final_der_CH}
\ee
In this expansion, the coefficient $A_2$ is identically zero. 

\begin{table}[htb]
\centering
\begin{tabular}{rll}
\hline\hline
\multicolumn{1}{c}{$\rho$}  & \multicolumn{1}{c}{${\cal A}^{\rm tri}_1(\rho)$}&
                              \multicolumn{1}{c}{$A_0^{\rm tri}(\rho)$} \\
\hline
1  &-16.556352382598901           & \phantom{-1}0.588715732188061 \\  
2  &-16.439945008735128           &           -12.618145549991986 \\
3  &\phantom{1}-6.161165303236093 &           -16.467144991961212 \\
4  &\phantom{-1}2.808024778142930 &           -15.425136698441144 \\
5  &\phantom{-1}6.829759193109778 &           -12.962673854292640 \\
6  &\phantom{-1}7.259238447062151 &           -10.685119155844771 \\
7  &\phantom{-1}6.078697042860241 & \phantom{1}-9.019533165133020 \\
8  &\phantom{-1}4.522637873985072 & \phantom{1}-7.925595102852174\\
9  &\phantom{-1}3.134987142065343 & \phantom{1}-7.248692542552664\\
10 &\phantom{-1}2.072903487004694 & \phantom{1}-6.845055620123084\\
11 &\phantom{-1}1.324966640033683 & \phantom{1}-6.610290784855592\\
12 &\phantom{-1}0.825436595929018 & \phantom{1}-6.476167809538218\\
13 &\phantom{-1}0.503936252639239 & \phantom{1}-6.400571208146704\\
14 &\phantom{-1}0.302641624812243 & \phantom{1}-6.358412464990734\\
15 &\phantom{-1}0.179285740783859 & \phantom{1}-6.335102866834869\\
16 &\phantom{-1}0.104987265970099 & \phantom{1}-6.322306687731851\\
17 &\phantom{-1}0.060870848500045 & \phantom{1}-6.315324296705190\\
18 &\phantom{-1}0.034988709413814 & \phantom{1}-6.311533950015708\\
19 &\phantom{-1}0.019959520236906 & \phantom{1}-6.309485615841503\\
20 &\phantom{-1}0.011309756764775 & \phantom{1}-6.308383036473573\\
$\infty$&\phantom{-1}0            & \phantom{1}-6.307116005647652\\
\hline\hline
\end{tabular}
\caption{\protect\label{table_f3_tri}
Values of the coefficients ${\cal A}^{\rm tri}_1(\rho)$ and 
$A_0^{\rm tri}(\rho)$ for several values of the torus aspect ratio $\rho$.
}
\end{table}

\begin{table}[htb]
\centering
\begin{tabular}{rll}
\hline\hline
\multicolumn{1}{c}{$\rho$}  & \multicolumn{1}{c}{${\cal A}^{\rm hc}_1(\rho)$}&
                              \multicolumn{1}{c}{$A_0^{\rm hc}(\rho)$} \\
\hline
1  &2.204248900857568 & \phantom{-}5.090947597599750 \\
2  &1.260776176148286 & \phantom{-}2.011554369485771 \\
3  &0.657422819248581 &           -0.855894359177661\\
4  &1.682404023579180 &           -2.151736456053733\\
5  &2.442900124268499 &           -2.489361833956631\\
6  &2.452406095152028 &           -2.473169556688759\\
7  &2.032154004962206 &           -2.378716655165567\\
8  &1.508563855204567 &           -2.293728702105723\\
9  &1.045164213119409 &           -2.234638841885926\\
10 &0.690994951708971 &           -2.197561506291189\\
11 &0.441659818805435 &           -2.175477057251966\\
12 &0.275146193270710 &           -2.162714473030565\\
13 &0.167978851785044 &           -2.155480460188504\\
14 &0.100880556816132 &           -2.151434956603961\\
15 &0.059761915864342 &           -2.149195097142150\\
16 &0.034995755659000 &           -2.147964640036792\\
17 &0.020290282882592 &           -2.147292993714472\\
18 &0.011662903145113 &           -2.146928331463238\\
19 &0.006653173413341 &           -2.146731247829992\\
20 &0.003769918921741 &           -2.146625156802025\\
$\infty$&0            &           -2.146503238450814\\
\hline\hline
\end{tabular}
\caption{\protect\label{table_f3_hc}
Values of the coefficients ${\cal A}^{\rm hc}_1(\rho)$ and
$A_0^{\rm hc}(\rho)$ for several values of the torus aspect ratio $\rho$.
}
\end{table}

The most important result contained in \reff{final_der_CH} is that the 
coefficient associated to the expected leading term $\sim L\log L$ vanishes. 
This is a highly non-trivial fact and we will discuss its physical meaning in 
Section~\ref{sec_irrel}. We have obtained the first four non-vanishing 
coefficients for the triangular lattice   
\begin{subeqnarray}
{\cal A}_1^{\rm tri}(\rho) &=& 2\rho E_1^{\rm tri}(\rho)  \left[ 
     \rho E_1^{\rm tri}(\rho)^2 + {36 \sqrt{3} \over \pi} \left( 
     {\sum |\theta_j| \real \log \theta_j \over \sum |\theta_j|} - 
     \log 2 |\eta| \right) \right] \\
A^{\rm tri}_{00} &=& - {216 \over \pi\sqrt{3} }  \\
A^{\rm tri}_0(\rho) &=& 48 - 54 d_0(\rho) + 6 \rho E_1^{\rm tri}(\rho)^2 \\
A^{\rm tri}_1(\rho) &=& 16 E_1^{\rm tri}(\rho) 
\label{final_A_tri}
\end{subeqnarray}
where the function $d_0(\rho)$ is defined in \reff{d0_final}.  
The numerical values of ${\cal A}_1^{\rm tri}$ and $A_0^{\rm tri}$ can
be found in Table~\ref{table_f3_tri}, while the numerical values of 
$A_1^{\rm tri}$ can be computed from Table~\ref{table_energy}.  
The coefficients for the hexagonal lattice are
\begin{subeqnarray}
{\cal A}_1^{\rm hc}(\rho) &=& 2\rho E_1^{\rm hc}(\rho)  \left[
     4 \rho E_1^{\rm hc}(\rho)^2 + {36 \over \pi\sqrt{3}} \left( 
     {\sum |\theta_j| \real \log \theta_j \over \sum |\theta_j|} - 
     \log 2 |\eta| \right) \right] \\
A^{\rm hc}_{00} &=& - {12 \over \pi }  \\
A^{\rm hc}_0(\rho) &=& {16\over3\sqrt{3}} - 3\sqrt{3} d_0(\rho) + 4\sqrt{3}
\rho E_1^{\rm hc}(\rho)^2 \\
A^{\rm hc}_1(\rho) &=& 4 E_1^{\rm hc}(\rho) 
\label{final_A_hc}
\end{subeqnarray}
and their numerical values can be found in Table~\ref{table_f3_hc}. 

\bigskip

\noindent
{\bf Remarks} 1. The coefficients \reff{final_A_tri}/\reff{final_A_hc} 
have the right behavior under the transformation $\rho\to 1/\rho$. 
In particular, they satisfy 
\begin{subeqnarray}
{\cal A}_1(\rho)  &=& \rho {\cal A}_1(1/\rho) \\
A_0(\rho)         &=& A_{00} \log \rho + A_0(1/\rho) \\
A_1(\rho)         &=& { A_1(1/\rho) \over \rho }
\end{subeqnarray}

2.- In the limit $\rho\to\infty$, both ${\cal A}_1(\rho)$ and $A_1(\rho)$ 
go to zero as in this limit $E_1(\rho)\to 0$ exponentially fast. The limit
of the coefficients $A_0(\rho)$ can be computed from 
(\ref{final_A_tri}c)/(\ref{final_A_hc}c) and \reff{lim_d0} 
\begin{subeqnarray}
\lim_{\rho\to\infty} A_0^{\rm tri}(\rho) &=& 48 - {72 \sqrt{3}\over\pi}
  \left( \gamma_E + \log {4\sqrt{3}\over \pi} \right) \\
\lim_{\rho\to\infty} A_0^{\rm hc}(\rho) &=& {16\over3\sqrt{3}} - {12\over \pi}
  \left( \gamma_E + \log {4\sqrt{3}\over \pi} \right) 
\end{subeqnarray}

3.- From Table~\ref{table_f3_tri} we see that ${\cal A}_1^{\rm tri}(\rho)$
vanishes at at value $\rho_{\rm min}$ between 3 and 4. Actually, 
${\cal A}_1^{\rm tri}$ is zero at 
\be 
\rho_{{\rm min},1}^{\rm tri}  \approx 3.6249264261
\ee
We also find that $A_0^{\rm tri}(\rho)$ vanishes at  
\be
\rho_{{\rm min},2}^{\rm tri} \approx 1.0300773853
\ee
In the hexagonal lattice, ${\cal A}_1^{\rm hc}(\rho)$ only vanishes in the 
limit $\rho\to\infty$, while $A_0^{\rm hc}$ is zero at 
\be
\rho_{{\rm min},2}^{\rm hc} \approx 2.6367691963
\ee

\subsection{Logarithmic finite-size corrections to $f^{(4)}_c$}
\label{sec_f4}

The full finite-size-scaling corrections to the fourth derivative of the 
free energy at criticality $f^{(4)}_c$ are rather cumbersome to compute. 
However, we can extract with much less effort the terms including logarithms. 
This is what we really need in the renormalization-group analysis of 
Section~\ref{sec_irrel}. 

The first step is the computation of the full expression for  $f^{(4)}_c$
in terms of the derivatives of the basic objects $Z_{\alpha,\beta}$. We 
should keep only those terms in which $Z_{\alpha,\beta}''(0)$, $Z_{00}'''(0)$ 
or $Z_{\alpha,\beta}^{(4)}(0)$ enter. There are three possible contributions
\begin{subeqnarray}
f^{(4)}_{c,\log,1} &=& {1\over V} 
         {\sum Z_{\alpha,\beta}''(0) \over \sum Z_{\alpha,\beta}(0) }  
         \left[ 4 \mu' \mu''' + 3 (\mu')^2 \right] 
\slabel{contribution_1} \\
f^{(4)}_{c,\log,2} &=& {6\over V} (\mu')^2 \mu'' 
         {Z_{00}'(0) \over \sum Z_{\alpha,\beta}(0) } 
         \left[  
         {Z_{00}'''(0) \over Z_{00}'(0) } -  
         3 {\sum Z_{\alpha,\beta}''(0) \over \sum Z_{\alpha,\beta}(0) } 
         \right] 
\slabel{contribution_2} \\
f^{(4)}_{c,\log,3} &=& {1\over V} (\mu')^4 
         \left[  
         {\sum Z_{\alpha,\beta}^{(4)}(0) \over \sum Z_{\alpha,\beta}(0) }
         -3 \left(
         {\sum Z_{\alpha,\beta}''(0) \over \sum Z_{\alpha,\beta}(0) }
         \right)^2  \right. \nonumber \\
         & & \qquad - 4 \left.  
         {Z_{00}'(0) \over \sum Z_{\alpha,\beta}(0) } 
         \left( 
         {Z_{00}'''(0) \over Z_{00}'(0) } -                          
         3 {\sum Z_{\alpha,\beta}''(0) \over \sum Z_{\alpha,\beta}(0) } 
         \right) 
         \right] 
\slabel{contribution_3}
\end{subeqnarray}
where the derivatives of $\mu$ with respect to $\beta$ have been represented
for short by $\mu'$, $\mu''$, etc. The first contribution 
\reff{contribution_1} is clearly non-zero and of order $\log N$. 
The second contribution \reff{contribution_2} does not actually contain 
any logarithm, as the logarithmic contributions of $Z_{00}'''(0)/Z_{00}'(0)$ 
and $-3\sum Z_{\alpha,\beta}(0)''/\sum Z_{\alpha,\beta}(0)$ cancel out exactly 
[See e.g., \reff{ratio_zpp_over_z_final}/\reff{final_ZPPP}]. The same 
argument applies to the second line of \reff{contribution_3}.

In order to compute the contribution of the first two terms in 
\reff{contribution_3}, we have to consider the fourth derivative of 
$Z_{\alpha,\beta}(\mu)$ at $\mu=0$ when $(\alpha,\beta)\neq (0,0)$. 
After some algebra, we find that the logarithmic contributions to that 
derivative are
\begin{eqnarray}
Z^{(4)}_{\alpha,\beta,\log}(0) &=& 
   {12 M N \over \pi\sqrt{3}} Z_{\alpha,\beta}''(0)  
   \left( {\cal S}^{(1)}_\alpha + 2 {\cal S}^{(2)}_{\alpha,\beta} \right) 
  + M^2 Z_{\alpha,\beta}''(0) \delta_{\alpha,0} \nonumber \\
 & & \qquad + 
  {8 M^3 N \over \pi\sqrt{3}} Z_{\alpha,\beta}(0) \log N \delta_{\alpha,0}
\label{def_Z4ab}
\end{eqnarray}
where the sums ${\cal S}^{(j)}$ are defined in \reff{def_s}. After some
more algebra we find that the contribution of \reff{contribution_3} does 
{\em not} contain any logarithms.  

In conclusion, we find that the finite-size-scaling expansion for the 
observable $f^{(4)}_c$ contains a {\em single} logarithmic term 
\be
f^{(4)}_{c,\log} = B_{00} \log N 
\ee
where $B_{00}$ can be read from \reff{contribution_1}. Its numerical value is
\be
B_{00} = \left\{ \begin{array}{ll}
                2736/(\pi\sqrt{3}) & \qquad \hbox{\rm triangular} \\[1mm]
                120/(\pi\sqrt{3})  & \qquad \hbox{\rm hexagonal}
                  \end{array} 
         \right.
\label{final_B00}
\ee
The leading term in the large-$N$ expansion of  $f^{(4)}_c$ is 
expected to be $\sim N^2$, and we also expect a term of order $\sim N$.   

%
%
\section{Irrelevant operators in the two-dimensional Ising model} 
\label{sec_irrel} 

Let us first collect the main results of the previous sections
\begin{subeqnarray}
\slabel{final_F}
f_c(N,\rho)     &=& f_{\rm bulk} + \sum\limits_{m=1}^\infty 
                 {f_{2m}(\rho) \over N^{2m} } \\
\slabel{final_E}
E_c(N,\rho)     &=& E_0 + \sum\limits_{m=0}^\infty 
                 {E_{2m+1}(\rho) \over N^{2m+1} } \\
\slabel{final_CH}
C_{H,c}(N,\rho) &=& C_{00}\log N + C_0(\rho) + \sum\limits_{m=1}^\infty
                 {C_{m}(\rho) \over N^m } \\
\slabel{final_F3}
f^{(3)}_c(N,\rho) &=& {\cal A}_1(\rho) N + A_{00}\log N + A_0(\rho) 
       + \sum\limits_{m=1}^\infty {A_{m}(\rho) \over N^m } \\
\slabel{final_F4}
f^{(4)}_{c,\log}(N,\rho) &=& B_{00}\log N 
\label{final_results}
\end{subeqnarray}
It is also important to mention that the coefficients $f_4$, $f_8$, $E_3$,
$E_7$, $C_2$, $C_3$, and $A_2$ vanish. 
In this Section we will use these results 
to study the irrelevant operators in the two-dimensional Ising model and 
the finite-size-scaling function $\widetilde{W}$ defined below.   
The results will be applicable to both
the triangular and hexagonal lattices as the analytic structure of the 
corresponding asymptotic expansions is the same. To our knowledge, there
are no predictions based on conformal field theory for the hexagonal-lattice
Ising model. In this section we will follow basically the notation of  
ref.~\cite{Caselle_01b}. 

Let us start with the basic finite-size-scaling Ansatz for a system defined on
a torus of linear size $L$ (the aspect ratio is also fixed and plays no role
in this discussion), zero magnetic field and reduced temperature $\tau$ 
\cite{Caselle_01b,Salas_Sokal_Ising}
\be
f(\tau;L) = f_b(\tau) + {1\over L^2} W(\{ \mu_j(\tau) L^{y_j}\}) + 
                        {\log L \over L^2} 
                        \widetilde{W}(\{ \mu_j(\tau) L^{y_j}\})  
\label{fss_ansatz}
\ee
where $f_b(\tau)$ is a regular function of $\tau$ and  
the scaling functions $W$ and $\widetilde{W}$ depend on the non-linear 
scaling fields $\mu_j(\tau)$ belonging to the identity and energy 
conformal families. Among them the only relevant field is the one associated
to the temperature $\mu_t(\tau)$ (See Table~\ref{table_oper}).  
In this Ansatz we have explicitly used the assumptions (a)-(c) 
introduced in Section~\ref{sec_intro}. 

The reduced temperature $\tau$ measures the distance to the critical 
point\footnote{ 
  This parameter should not be confused with the torus modular parameter. 
  In this section $\tau$ will mean the reduced temperature, while in the 
  rest of the paper it will denote the usual modular parameter.
}
and it is defined such that $\tau=0$ at $\beta=\beta_c$ and 
$\tau>0$ (resp.\  $\tau<0$) for $\beta<\beta_c$ (resp.\   $\beta>\beta_c$).   
In the Ising model on the triangular and hexagonal lattices this 
parameter takes the form
\be
\tau = \left\{ \begin{array}{ll} 
  {1 + v^2 - 4v \over (1-v) \sqrt{2v}} & \qquad \qquad \hbox{\rm triangular}
   \\[2mm]
  {1-3v^2 \over v\sqrt{2(1-v^2)}}      & \qquad \qquad \hbox{\rm hexagonal}
               \end{array}\right.
\label{def_tau}
\ee
where as usual $v=\tanh \beta$. 
Under the transformations that map the high-temperature phase onto the 
low-temperature phase and viceversa 
\be
v \to v' = \left\{ \begin{array}{ll} 
  \left( {\sqrt{1-v+v^2} -\sqrt{v} \over 1-v} \right)^2 
               & \qquad \qquad \hbox{\rm triangular}\\[2mm]
  \sqrt{1-v^2 \over 1+3v^2} 
               & \qquad \qquad \hbox{\rm hexagonal}
               \end{array}\right.
\label{def_vprime}
\ee
the reduced temperature simply maps as $\tau\to -\tau$. Equations 
\reff{def_tau}/\reff{def_vprime} in the triangular-lattice case were 
introduced in \cite{Caselle_01b}.  

The non-linear scaling fields $\mu_j(\tau)$ can be written as a power series 
in $\tau$
\be
\mu_j(\tau) = \mu_j(0) + \tau \mu_{1,j} + {1\over 2}\tau^2 \mu_{2,j} + \cdots
\label{def_scaling_fields}
\ee
and we usually take the normalization $\mu_j(0)=1$ for the identity-family
fields, and $\mu_{1,j}=1$ for the energy-family fields (These latter scaling
fields are odd under the transformation $\tau\to-\tau$, thus they satisfy
$\mu_j(0)=0$).   

As explained in ref.~\cite{Caselle_01b}, both scaling functions satisfy 
\be
\label{eq_scaling_functions}
W(\{ \mu_j(-\tau) (-L)^{y_j} \}) = W(\{ \mu_j(\tau) L^{y_j} \}) 
\ee
(and analogously for $\widetilde{W}$). Thus, even (resp.\  odd) derivatives 
of $W$ and $\widetilde{W}$ with respect to $\tau$ will contain only 
even (resp.\  odd) powers of $L$. This fact explains the structure found for
the internal-energy and specific-heat expansions: 
\begin{subeqnarray}
-E_c(L) &=& \left. {\partial \tau \over \partial \beta} \right|_{\beta=\beta_c} 
            \, 
            \left. {\partial f \over \partial \tau} \right|_{\tau=0} \\
C_{H,c}(L) &=& 
        \left. {\partial^2 \tau \over \partial \beta^2} \right|_{\beta=\beta_c}
        \, 
        \left. {\partial f \over \partial \tau} \right|_{\tau=0} +
        \left. {\partial \tau \over \partial \beta} \right|_{\beta=\beta_c}
        \, 
        \left. {\partial f^2 \over \partial \tau^2} \right|_{\tau=0}
\label{equations_1}
\end{subeqnarray}  
In particular, \reff{equations_1} shows why the odd powers of the specific-heat
expansion are related to those of the internal energy. 
We will also make the following assumption, which is motivated by the 
absence of terms $L^{-m} \log L$ for any $m>0$ in the expansions 
\reff{final_results} 
\begin{itemize}
\item[(d)] The scaling function $\widetilde{W}$ only depends on the scaling
field associated to the temperature 
\be
\widetilde{W}(\{\mu_j(\tau) L^{y_j}\}) = \widehat{W} (\mu_t(\tau)L)
\ee
\end{itemize}
As we know that there are no logarithmic contributions to the free and 
internal energies \reff{final_F}/\reff{final_E}, the scaling function 
$\widehat{W}$ should satisfy 
\be
\widehat{W}(0)  = \left. {\partial \widehat{W}(x) \over \partial x}
                  \right|_{x=0} = 0 
\ee

Conformal field theory \cite{Caselle_01b} provides a list of irrelevant 
operators that may appear in the two-dimensional Ising model 
(See Table~\ref{table_oper}). By comparing the finite-size-scaling Ans\"atze
for the free energy, internal energy and specific heat obtained from 
\reff{fss_ansatz} to the corresponding exact results \reff{final_results} we
may get new insights about the operator content of the model.

\begin{table}[htb]
\centering
\begin{tabular}{|c|l|l|rr|}
\hline\hline
\multicolumn{1}{|c|}{Family}  & 
\multicolumn{1}{|c|}{$j$} &
\multicolumn{1}{|c|}{$\mu_j$} &
\multicolumn{1}{c}{$s$} & 
\multicolumn{1}{c|}{$y$} \\
\hline
$[I]$ & 0 & $I$                  & 0 &  2 \\ 
      & 1 & $T\bar{T}$           & 0 & -2 \\
      & 2 & $T^3 + \bar{T}^3$    & 6 & -4 \\
      & 3 & $Q_4^I \bar{Q}_4^I$  & 0 & -6 \\
      & 4 & $\bar{Q}_2^I Q_8^I + Q_2^I \bar{Q}_8^I$  
                                 & 6 & -8 \\
      & 5 & $Q_{12}^I +\bar{Q}_{12}^I$  
                                 & 12 & -10 \\
\hline 
$[\epsilon]$ 
      &$t$& $\epsilon$           & 0  & 1  \\
      & 6 & $Q_6^\epsilon + \bar{Q}_6^\epsilon$ 
                                 & 6  & -5 \\
      & 7 & $Q_4^\epsilon\bar{Q}_4^\epsilon$ 
                                 & 0  & -7 \\
\hline\hline
\end{tabular}
\caption{\protect\label{table_oper}
Operators in the two-dimensional Ising model according to 
ref.~\protect\cite{Caselle_01b}. For each conformal family, we have listed 
the primary and quasiprimary fields belonging to it. For each scaling 
field $\mu_j$, we show the notation used in ref.~\protect\cite{Caselle_01b},
its spin $s$ and its renormalization-group exponent $y$. We have included 
the most relevant fields (i.e., $y\geq -10$) with spin $s = 6\, {\mathbb N}$.
We have omitted the conformal family $[\sigma]$ as it is irrelevant in this
discussion. Only the primary fields $I$ and $\epsilon$ are relevant.  
}
\end{table}

Let us start with the free energy. At the critical point $\tau=0$ this can be 
written as  
\be
f_c(L) = f_b(0) + {1\over L^2} W(\{ x_j\} ) 
\label{fss_F}
\ee
where $W$ depends only on the identity-family fields through the variables
$x_j = \mu_j(0) L^{y_j}$, as the energy-family scaling fields vanish at 
criticality. This expression can be Taylor expanded for large $L$, so we 
obtain a power series in $L^{-1}$. The exact result \reff{final_F} tells us 
that only corrections of order $L^{2m}$ can occur, except for the terms 
of order $L^{-4}$ and $L^{-8}$. From Table~\ref{table_oper}, we see that 
the scaling fields $T\bar{T}$ and $\mu_3$ precisely give corrections 
of order $L^{-4}$ and $L^{-8}$ to the expansion of \reff{fss_F}. Hence, 
we need to impose the conditions 
\be
\mu_{T\bar{T}}(0) = \mu_3(0) = 0
\label{res_oper_1}
\ee

The derivative of the free energy with respect to $\tau$ can be written as
\cite{Caselle_01b}
\begin{subeqnarray}
\left. {\partial f \over \partial\tau}  \right|_{\tau=0} &=& 
\left. {\partial f_b \over \partial\tau}\right|_{\tau=0} + 
{1\over L^2} \sum\limits_{j\in [\epsilon]} L^{y_j} \mu_{1,j} W_j(\{ x_k\})  
\slabel{fss_e} \\ 
   &=& \left. {\partial f_b \over \partial\tau}\right|_{\tau=0} + 
{1\over L} \mu_{1,t} W_t(\{ x_k\}) + {1\over L^7} \mu_{1,6} W_6(\{ x_k\}) 
       \nonumber \\
   & & \qquad +{1\over L^9} \mu_{1,7} W_7(\{ x_k\}) + \cdots 
\end{subeqnarray}
Each function $W_j(\{ x_k\})$ can be expanded as we did for the free energy, 
giving a power series in $L^{-2m}$ with no contribution to orders $L^{-4}$ and 
$L^{-8}$.  From the exact solution \reff{final_E}, we 
see that only corrections of the type $L^{-2m-1}$ can appear except for the
powers $L^{-3}$ and $L^{-7}$. This implies  
that the scaling field $\mu_6$ cannot play any role, thus 
\be
\mu_{1,6} = 0
\label{res_oper_2}
\ee

The second derivative of the free energy at criticality is given by 
\cite{Caselle_01b}: 
\begin{eqnarray}
\left. {\partial^2 f   \over \partial\tau^2}\right|_{\tau=0} &=&
\left. {\partial^2 f_b \over \partial\tau^2}\right|_{\tau=0} +
{1\over L^2} \sum\limits_{i,j\in [\epsilon]} L^{y_i+y_j} W_{ij}(\{ x_k\})
+ {1\over L^2} \sum\limits_{j\in [I]} \mu_{2,j} L^{y_j} W_j(\{ x_k\}) 
\nonumber \\ 
 & & \qquad + 2 \log L 
\left. {\partial^2 \widehat{W}(x) \over \partial x^2}\right|_{x=0}
\slabel{fss_ch} 
\end{eqnarray}
where we have used the standard normalization.  
The second term in the r.h.s. of \reff{fss_ch} can be written as
\be
{1\over L^2} \sum\limits_{i,j\in [\epsilon]} L^{y_i+y_j} W_{ij}(\{ x_k\}) 
= W_{tt}(\{ x_k\}) + {1\over L^6} W_{t7}(\{ x_k\}) + \cdots 
\ee
These two terms alone give all even powers of $L^{-1}$ except $L^{-2}$ in 
agreement with the exact expansion \reff{final_CH}. The third term in the 
r.h.s. of \reff{fss_ch} is equal to
\be
{1\over L^2} \sum\limits_{j\in [I]} \mu_{2,j} L^{y_j} W_j(\{ x_k\})
   = {1\over L^4} \mu_{2,1} W_1(\{ x_k\})
   + {1\over L^6} \mu_{2,2} W_2(\{ x_k\}) + \cdots 
\ee
Again, this a power series containing all even powers of $L^{-1}$ except 
$L^{-2}$ in agreement with \reff{final_CH}. 

The coefficient of the leading term should be equal to $C_{00}$
\be
C_{00} = \left( \left. {d\tau \over d\beta}\right|_{\beta=\beta_c}\right)^2 
\widehat{W}''(0) 
\ee
Hence we can determine the numerical value of $\widehat{W}''(0)$ by using
(\ref{C_tri_final}a)/(\ref{C_hc_final}a) and the definition of $\tau$
\reff{def_tau}. The result is 
\be
\widehat{W}''(0) = \left\{ \begin{array}{ll}
       1/ ( \pi\sqrt{3}) & \qquad \qquad \hbox{\rm triangular} \\[2mm]  
       1/ (2\pi\sqrt{3}) & \qquad \qquad \hbox{\rm hexagonal} 
       \end{array}\right. 
\label{final_wpp0}
\ee
where we have considered the standard normalization for $\mu_t(\tau)$. The
value \reff{final_wpp0} for the triangular lattice agrees with the result
obtained in \cite[eq.~(3.34)]{Caselle_01b}. 

\bigskip

\noindent
{\bf Remarks}. 1. The irrelevant scaling fields belonging to the identity 
   family that may play a role have non-zero spin (namely, $s=6,12$).  
   The spin-zero fields belonging to this family should vanish at criticality 
   (e.g., $\mu_j(0) =0$). This result agrees with 
   Conjecture~\ref{conjecture_Caselle}. 

2. The vanishing of the field $\mu_1=T\bar{T}$ at criticality 
   supports Conjecture~(d0) of \cite{Caselle_01b}. However, our results 
   do not imply their stronger Conjecture~(d1): the scaling field $T\bar{T}$ 
   decouples (i.e., $\mu_{T\bar{T}}(\tau) = 0$ for all $\tau$).\footnote{
   It is worth mentioning that the authors of \protect\cite{Caselle_01b} 
   showed by considering the large-distance behavior of the triangular-lattice 
   Ising model two-point function that  
   $\mu_{T\bar{T}}(\tau) = o(\tau^4)$. This result strongly supports 
   their Conjecture~(d1).   
}

3. In the internal-energy analysis, we concluded that the irrelevant field
   $\mu_6$ should vanish at criticality \reff{res_oper_2}. This operator has
   spin six, therefore this result is {\em not} implied by 
   Conjecture~\ref{conjecture_Caselle}. In other words, there are 
   cancellations also in the non-scalar sector.  
   On the other hand, we find no constraint on the spin-zero 
   irrelevant field $\mu_7$. However, if Conjecture~\ref{conjecture_Caselle}
   is true, then we should have $\mu_{1,7}=0$. 

4. In order to obtain the exact solutions \reff{final_results}  we need to 
   include {\em at least} two irrelevant operators. This result agrees with
   the findings of \cite{Orrick_00a,Orrick_00b} for the square-lattice 
   model. 
   It is worth noticing that we can formally obtain the exact solutions 
   \reff{final_results} by including the spin-6 irrelevant scaling field
   $\mu_2 = T^3+\bar{T}^3$ with $y=-4$ and the spin-12 field $\mu_5$ with 
   $y=-10$. 
\begin{subeqnarray}
f_c(L) &=& f_b(0) + {1\over L^2} W(\mu_2(0) L^{-4},\mu_5(0) L^{-10}) \\
\left. {\partial f \over \partial\tau}  \right|_{\tau=0} &=&
\left. {\partial f_b \over \partial\tau}\right|_{\tau=0} +
       {1\over L} W_t(\mu_2(0) L^{-4},\mu_5(0) L^{-10} ) \\
\left. {\partial^2 f   \over \partial\tau^2}\right|_{\tau=0} &=&
2 \widehat{W}''(0) \log L +  
\left. {\partial^2 f_b \over \partial\tau^2}\right|_{\tau=0} 
       + W_{tt}(\mu_2(0) L^{-4},\mu_5(0) L^{-10}) 
  \nonumber \\
 & & \qquad + {1\over L^6} \mu_{2,2} W_2(\mu_2(0) L^{-4},\mu_5(0) L^{-10}) 
  \nonumber \\ 
 & & \qquad + {1\over L^{12}} \mu_{2,5} W_5(\mu_2(0) L^{-4},\mu_5(0) L^{-10})
\end{subeqnarray}

\bigskip 

Let us now consider the observable $f^{(3)}_c$ \reff{def_der_CH}.  
We are interested here only in the terms containing logarithms, 
which are directly related to derivatives of the scaling function 
$\widehat{W}(x)$. The contribution of this scaling function to this 
observable can be written as
\be
f^{(3)}_{c,\log} =  
L \log L \; \widehat{W}'''(0) 
\left({\partial \tau \over \partial\beta}\right)_{\beta=\beta_c}^3 + 
3 \log L \; \widehat{W}''(0)
\left.{\partial \tau \over \partial\beta}\right|_{\beta=\beta_c}  
\left.{\partial^2 \tau \over \partial\beta^2}\right|_{\beta=\beta_c} 
\ee
The exact result \reff{final_der_CH} shows that 
\be
\left. {\partial^3 \widehat{W}(x) \over \partial x^3 }\right|_{x=0} = 0
\ee
This result is consistent with the conjecture put forth by the authors of 
ref.~\cite{Caselle_01b} who claimed that the scaling function $\widehat{W}$ 
is quadratic in is argument (i.e., $\widehat{W}(x) = A x^2$). 

On the other hand, the coefficient of the logarithmic term in  
$f^{(3)}_c$ (i.e., $A_{00}$) is proportional to $\widehat{W}''(0)$. 
This observation provides another way to compute the quantity 
$\widehat{W}''(0)$ and a direct mean to test the predictions \reff{final_wpp0}.
By using the exact results 
(\ref{final_A_tri}b)/(\ref{final_A_hc}b)/\reff{def_tau}, we arrive at the
same values as in \reff{final_wpp0} supporting the correctness of our 
results.   

Finally, we will discuss the observable $f^{(4)}_c$. The contribution
of the scaling function $\widehat{W}$ to this observable is given by
\begin{eqnarray}
f^{(4)}_{c,\log} &=&  
L^2 \log L \; \widehat{W}^{(4)}(0)
\left({\partial \tau \over \partial\beta}\right)
 \nonumber \\ 
 & & \quad  +
3 \log L \; \widehat{W}''(0) \left[ 
 4 {\partial \tau \over \partial\beta}{\partial^3 \tau \over \partial\beta^3}
+3 \left( {\partial^2 \tau \over \partial^2\beta}\right)^2 + 
 4  \left( {\partial^2 \tau \over \partial^2\beta}\right)^4 \mu_{3,t} 
\right] 
\label{Ansatz_log_F4}
\end{eqnarray}
where all the derivatives of $\tau$ with respect to $\beta$ should be 
evaluated at $\beta=\beta_c$. By comparing the above formula to 
\reff{final_F4}/\reff{final_B00}, we conclude that
\be
\left. {\partial^4 \widehat{W}(x) \over \partial x^4 }\right|_{x=0} = 0
\ee
This result is compatible with $\widehat{W}(x)$ being a quadratic function 
of $x$. 
On the other hand, as we know the numerical values of the derivatives of 
$\tau$ w.r.t. $\beta$ for the triangular and hexagonal lattices, we can
use equations~\reff{final_B00}/\reff{Ansatz_log_F4} to deduce the value of
$\mu_{3,t}$. The result is the same for both lattices $\mu_{3,t}=-1/4$, so 
the non-linear scaling field $\mu_t$ depends on $\tau$ in the following way
\be
\mu_t(\tau) = \tau - {1\over 24} \tau^3 + {\cal O}(\tau^5) 
\label{final_mut}
\ee
This relation coincides with the function $a(\tau)$ obtained in 
ref.~\cite{Caselle_01b} for the triangular lattice:\footnote{
  It is not hard to realize that the function $a(\tau)$ 
  \protect\reff{exact_a_tri} is the same for the 
  hexagonal lattice. The key observation is that the free energy for this 
  lattice in the thermodynamic limit \protect\reff{f_bulk_hc_infty} 
  can be written as
$$
  f_{\rm bulk}^{\rm hc} = {1\over4}\int_0^\pi \int_0^\pi \, 
    {dx dy \over 4\pi^2} \, 
    \log \left[ 3 + \tau^2 - \omega(x,y) \right] + \hbox{\rm cnt.} 
$$
  where $\tau$ is given by \protect\reff{def_tau}. This equation is equivalent 
  to the definition used in \protect\cite{Caselle_01b} to define $a(\tau)$ 
  for the triangular lattice.
}
\be 
a(\tau) = \tau - {1\over 24} \tau^3 + {47\over 10368} \tau^5 - 
  {161\over 248832}\tau^7 + {\cal O}(\tau^9) 
\label{exact_a_tri}
\ee
The equality between $a(\tau)$ and $\mu_t(\tau)$ is important because  
it provides support to Conjecture~\ref{conjecture_Caselle}: 
if this conjecture is correct, then both function should
coincide \cite{Caselle_01b}.

We can summarize the results obtained on the scaling function $\widehat{W}$ 
in the following conjecture (which is a natural extension of the conjecture 
$\widehat{W}(x) = x^2/(2\pi)$ for the 
square-lattice model \cite{Caselle_01b}):
\begin{conjecture} \label{conj_w}
In the Ising model on the triangular and hexagonal lattices with toroidal 
boundary conditions, the scaling function $\widehat{W}$ is a function solely 
of the argument $x=\mu_t(\tau) L$ and this function is equal to
\be
 \widehat{W}(x) = \left\{ \begin{array}{ll}
          x^2/(2\pi\sqrt{3}) & 
                         \qquad \qquad \hbox{\rm{\em triangular}}\\[3mm] 
          x^2/(4\pi\sqrt{3}) & 
                         \qquad \qquad \hbox{\rm{\em hexagonal}} 
                         \end{array} \right. 
\ee
\end{conjecture}

The coefficient of $\widehat{W}$ should coincide with the constant $A$
obtained in the infinite-volume limit analysis of the triangular-lattice 
model \cite[Eq.~(2.34)]{Caselle_01b}. The agreement between those coefficients 
adds support to this conjecture. 

%
%
\section{Further remarks and conclusions} \label{sec_conclusions}

We have obtained the asymptotic expansions for the free energy, internal 
energy, specific heat and $f^{(3)}$ of a critical Ising model on the 
triangular and hexagonal lattices wrapped on a torus of width 
$N$ and aspect ratio $\rho$. These expansions are given in 
\reff{final_results}.  
In particular, we have found the exact coefficients $f_{\rm bulk}$, $f_2$,
$f_4 = f_8 = 0$, $f_6$, $E_0$, $E_1$, $E_3=E_7=0$, $E_5$, $C_{00}$, $C_0$, 
$C_1$, $C_2=C_3=0$, $C_4$, $C_5$, ${\cal A}_1$, $A_{00}$, $A_0$, $A_1$, 
$A_2=0$, and $B_{00}$ for both lattices.  

The first important observation is that the analytic structure of the 
finite-size corrections of the observables considered in this paper is the 
same for the triangular- and the  hexagonal-lattice models. The reason of
this coincidence is that both lattices have the same underlying 
Bravais lattice. This agrees with the physical content of 
Conjecture~\ref{conjecture_Caselle}: as they have the same rotational 
symmetry group, they should have the same irrelevant operators, hence
leading to the same finite-size corrections.  

As it can be seen in \reff{final_results}, all 
the corrections are integer powers of $N^{-1}$. The only exceptions are the
logarithmic terms in the specific heat \reff{final_CH}, 
$f^{(3)}_c$ \reff{final_F3}, and $f^{(4)}_c$ \reff{final_F4}. 
In the first case this term is the leading one,
while in the other ones it is subleading.     
In the free-energy expansion \reff{final_F} only even powers of $N^{-1}$ 
can occur, while in the internal-energy expansion \reff{final_E} only 
odd powers of $N^{-1}$ appear. In the specific-heat expansion even and
odd powers of $N^{-1}$ occur. Furthermore, the odd coefficients in this
latter expansion are proportional to the corresponding odd coefficients in 
the internal-energy expansion. The constant depends on how the 
mass $\mu$ \reff{def_mass} depends on the temperature \reff{proportional_law}.  
In the expansion of the observable $f^{(3)}$, we find corrections with all 
powers of $N^{-1}$ except for the term $N^{-2}$.
Indeed, the coefficients $f_m$, $E_m$, $C_m$ and $A_m$ do depend on the 
lattice structure of the model, hence they are not universal.

The fact that $E_{2m+1}/C_{2m+1}$ is a $\rho$-independent number for the square
lattice ($=-1/\sqrt{2}$) \cite{Salas_01,Izmailian2}, 
suggested the idea that this ratio might be 
universal (e.g., it does {\em not} depend on the lattice structure).\footnote{ 
   Izmailian and Hu \protect\cite{Izmailian3} 
   (see also \protect\cite{Okabe_01}) computed the finite-size expansion of 
   the free energy $f(N)=f_{\rm bulk} + \sum_{k=1}^\infty f_k/N^{2k}$ 
   and the inverse correlation length 
   $\xi^{-1}(N) = \sum_{k=1}^\infty b_k/N^{2k-1}$ for a 
   critical Ising model 
   on several $N\times\infty$ lattices (i.e., square, hexagonal and triangular)
   with periodic boundary conditions. They found lattice-dependent coefficients
   $f_k$ and $b_k$, but universal ratios $b_k/f_k = (2^{2k}-1)/(2^{2k-1}-1)$. 
}
However, our results show that this is not the case: 
\be
{E_{2m+1}(\rho) \over C_{2m+1}(\rho) } = \left\{ \begin{array}{ll}
                  -1/2        & \qquad \hbox{\rm triangular} \\ 
                  -\sqrt{3}/2 & \qquad \hbox{\rm hexagonal}  
                  \end{array}
                  \right.
\label{proportional_tri_hc}
\ee
Thus, the proportional constant {\em does} depend on the lattice structure,
hence it is not universal. We can write \reff{proportional_tri_hc} and 
the corresponding square-lattice relation \reff{ratio_sq} in an 
unified way by realizing that the constant is just $-1/E_0$. 

It is important to note\footnote{We thank Andrea Pelissetto for useful 
 comments on this matter.
}
that one key ingredient in this discussion is the fact that there is an 
exact transformation \reff{def_vprime} mapping the high-temperature
phase onto the low-temperature phase, so we can define a parameter $\tau$ 
\reff{def_tau} transforming as $\tau\to-\tau$. It is not clear whether this
transformation exists or not for an Ising model defined on a general lattice.
However, if that transformation does exist, then we can define $\tau$ 
so eq.~\reff{eq_scaling_functions} holds, leading 
to \reff{equations_1}/\reff{proportional_tri_hc}.  
We can summarize all these observations in the following conjecture:

\begin{conjecture} \label{conjecture_ratio}
Let us consider a critical Ising system on a regular two-dimensional lattice
with toroidal boundary conditions. Let us further assume that there is 
an exact mapping $v\to v'$ from the high-temperature phase onto the 
low-temperature phase such that $\tau\to -\tau$. Then, the internal energy 
and specific heat can be expanded in power series of $N^{-1}$ as in  
\protect\reff{final_E}/\protect\reff{final_CH} 
and the coefficients $E_m(\rho)$ and $C_m(\rho)$ satisfy   
\be
{E_{m}(\rho) \over C_m(\rho) } = \left\{ \begin{array}{ll}
                  -1/E_0       & \qquad \hbox{\rm{\em for $m$ odd}} \\
                  \phantom{-}0 & \qquad \hbox{\rm{\em for $m$ even}}
                  \end{array}
                  \right.
\label{ratio_general}
\ee
where $E_0$ is the bulk internal energy [See \protect\reff{final_E}]. Indeed, 
we understand that this ratio is not defined whenever $E_m=0$. 
\end{conjecture}

If this conjecture is true, then we could define the expansions
\begin{subeqnarray}
\slabel{final_E_bis}
E_c(N,\rho)     &=& E_0\left[ 1  + \sum\limits_{m=0}^\infty
                 {\widetilde{E}_{2m+1}(\rho) \over N^{2m+1} } \right] \\
\slabel{final_CH_bis}
C_{H,c}(N,\rho) &=& E_0^2 \left[ \widetilde{C}_{00}\log N + 
                                 \widetilde{C}_0(\rho) + 
                 \sum\limits_{m=1}^\infty
                 {\widetilde{C}_{m}(\rho) \over N^m }  \right]
\label{final_results_bis}
\end{subeqnarray} 
and then the new ratios would be universal 
\be
{\widetilde{E}_{m}(\rho) \over \widetilde{C}_m(\rho) } = 
                        \left\{ \begin{array}{ll}
                  -1           & \qquad \hbox{\rm for $m$ odd} \\
                  \phantom{-}0 & \qquad \hbox{\rm for $m$ even}
                  \end{array}
                  \right.
\label{ratio_general_bis}
\ee

In Section~\ref{sec_intro} we mentioned that the results contained in this 
paper could serve also to test Monte Carlo simulations. Indeed, the 
expressions \reff{final_Zab}/\reff{def_Z00p}/\reff{def_ZPP0} provide a 
way to compute the {\em exact} values of the internal energy and 
specific heat for any finite torus of size $N\times M$. For very large
lattices one could also use the (easier to evaluate) asymptotic 
expansions \reff{final_E}/\reff{final_CH}.  

On the other hand, by taking the {\em exact} values of any observable 
for fixed aspect ratio $\rho$ and several values of the torus width $N$, we 
can check whether the asymptotic expansions \reff{final_results} are 
correct or not. In particular, by fitting the exact values to the 
corresponding Ansatz, we can verify whether the numerical coefficients 
coincide with the estimates coming from the fits.  We have performed such 
analysis and we have confirmed that the numerical values of the coefficients 
$f_m$, $E_m$, $C_m$ and $A_m$ for several values of $\rho$ coincide with
the estimates coming from the fits. In addition, this procedure allows us
to obtain crude estimates of the next coefficients in each expansion. For 
instance, we obtain for $\rho=1$ (which is the case most frequently 
considered in the literature) the following values  
\begin{subeqnarray}
f_{10}^{\rm tri}(1)   &\approx&   \phantom{-}1.932 \;, \qquad 
f_{10}^{\rm hc}(1)\,\; \approx\;  \phantom{-}0.966 \\ 
E_9^{\rm tri}(1)      &\approx&             -7.821 \;, \qquad 
E_9^{\rm hc}(1)     \; \approx\;            -2.258 \\ 
C_6^{\rm tri}(1)      &\approx&             -0.722 \;, \qquad
C_6^{\rm hc}(1)     \; \approx\;            -0.120 \\ 
A_3^{\rm tri}(1)      &\approx&   \phantom{-}9.124 \;, \qquad 
A_3^{\rm hc}(1)     \; \approx\;  \phantom{-}0.878
\end{subeqnarray}

\appendix
%
%
\section{The Euler-MacLaurin formula} \label{sec_Euler}

The Euler-MacLaurin formula is one important tool we need to compute 
asymptotic series. Here we will use the version of 
ref.~\cite[formula 23.1.32]{Abramowitz}. Let $F(x)$ be a function whose 
first $2n$ derivatives are continuous in the interval $(a,b)$. If we  
divide the interval into $m$ equal parts (so that $h=(b-a)/m$), then we
have  
\begin{eqnarray}
& & \sum\limits_{k=0}^{m-1} F(a + k h + \alpha h) = 
            {1\over h} \int_a^b F(t) dt \nonumber \\
  & & \qquad \qquad \qquad 
+ \sum\limits_{k=1}^p {h^{k-1} \over k!} B_{k}(\alpha) 
   [F^{(k-1)}(b) - F^{(k-1)}(a)]
  \nonumber \\
  & & \qquad  \qquad \qquad - {h^p \over p!} 
   \int_0^1 \widehat{B}_p(\alpha-t) \left\{ \sum\limits_{k=0}^{m-1} 
    F^{(p)}(a + k h + t h)  \right\} dt  
\label{Euler_MacLaurin_general}
\end{eqnarray}
where $p \leq 2n$, $0\leq \alpha \leq 1$,   
$\widehat{B}_n(x) = B_n(x-\lfloor x\rfloor)$ and $B_n(x)$ are the 
Bernoulli polynomials defined in terms of the Bernoulli numbers $B_k$ by 
\be
B_n(x) = \sum\limits_{k=0}^n \left(\begin{array}{l}
                                     n \\
                                     k
                                   \end{array} \right) B_k x^{n-k}  
\label{def_Bna}
\ee
Indeed, $B_n(0) = B_n$. The Bernoulli polynomials satisfy the identity
\cite[eq.~23.1.21]{Abramowitz}:
\be
B_n(1/2) = \left( {1\over 2^{n-1}} - 1 \right) B_n  
\label{B_onehalf}
\ee

We are mainly interested in sums of the form
\be
{1 \over L} \sum\limits_{n=0}^{\gamma L -1 } F\left( {2\pi\over L}(n+\alpha) 
            \right)  
\label{goal_sum}
\ee
The asymptotic expansion of the sum \reff{goal_sum}
in the limit $L\rightarrow\infty$ with $\gamma$ fixed can be obtained from 
\reff{Euler_MacLaurin_general}. If we assume that all the derivatives of 
$F(t)$ are integrable over the interval 
$[0,2\pi\gamma]$ we can formally
extend the sum in \reff{Euler_MacLaurin_general} to $k=\infty$ and drop
the remainder term [namely, the integral in \reff{Euler_MacLaurin_general}].  
In this case we can write the Euler-MacLaurin formula as follows 
\begin{eqnarray}
& & {1 \over L} \sum\limits_{n=0}^{\gamma L-1} F\left( {2\pi\over L}(n+\alpha)
   \right)  
 = {1\over 2\pi} \int_0^{2\pi\gamma} F(t) \, dt \nonumber \\
  & & \qquad \qquad \qquad  
+ {1\over 2\pi} \sum\limits_{k=1}^\infty \left( {2\pi \over L}\right)^k 
  {B_{k}(\alpha) \over k!} \left[
  F^{(k-1)}(2\pi\gamma) - F^{(k-1)}(0) \right]
\label{Euler_MacLaurin_formula}
\end{eqnarray}
In this paper we need the above formula in the particular case 
$L=2N$ and $\gamma=1/2$. Then \reff{Euler_MacLaurin_formula} reads  
\begin{eqnarray}
& & {1 \over N} \sum\limits_{n=0}^{N-1} F\left( {\pi\over N}(n+\alpha)
   \right)
 = {1\over \pi} \int_0^{\pi} F(t) \, dt \nonumber \\ 
 & & \qquad \qquad\qquad  
  + {1\over \pi} \sum\limits_{k=1}^\infty \left( {\pi \over N}\right)^k
  {B_{k}(\alpha) \over k!} \left[
  F^{(k-1)}(\pi) - F^{(k-1)}(0) \right]
\label{Euler_MacLaurin_formula_final}
\end{eqnarray} 

In the computation of the specific heat we also need formula 
\reff{Euler_MacLaurin_general} in the particular case $\alpha=0$ and $h=1$. 
In this case we can formally write \reff{Euler_MacLaurin_general} for a 
function $F$ whose derivatives are all integrable 
over $[a,b]$ in the following form \cite{Caracciolo_98} 
\begin{eqnarray}
\sum\limits_{k=a}^{b-1} F(k) &=& \int_a^b F(t) dt 
 - {1\over 2} \left[ F(b) - F(a) \right] \nonumber \\
  & & \qquad 
+ \sum\limits_{k=1}^\infty {B_{2k} \over (2k)!} [F^{(2k-1)}(b) - F^{(2k-1)}(a)]
\label{Euler_MacLaurin_bis}
\end{eqnarray}
where we have used the fact that \cite{Abramowitz}
\be
B_{2k+1} = \left\{ \begin{array}{ll}
                    -{1\over 2}  & \qquad k=0 \\
                    \phantom{-}0 & \qquad k>0 
                   \end{array}
                   \right.  
\ee
As we did in \cite{Salas_01}, we can apply \reff{Euler_MacLaurin_bis} to the
function $F(x) = x^{2m}$ with $a=0$ and $b=1$. We then obtain the identity
\be
\sum\limits_{k=1}^m {B_{2k}\over 2k} \left( \begin{array}{c}  
                                            2m \\
                                            2k-1 
                                            \end{array} \right) = 
{1\over 2} - {1\over 2m+1} 
\label{relation1}
\ee
If we apply \reff{Euler_MacLaurin_bis} to the case 
$F(x)=x^{2m-1}$ with the same endpoints as before, we obtain 
\be
\sum\limits_{k=1}^{m-1} {B_{2k}\over 2k} \left( \begin{array}{c} 
                                            2m-1 \\
                                            2k-1 
                                            \end{array} \right) = 
{1\over 2}\left( 1  - {1\over m} \right)  
\label{relation2}
\ee
%

%
%
\section{Theta functions} \label{sec_theta}

In this appendix we gather all the definitions and properties of the 
Jacobi's $\theta$-functions needed in this paper. We first introduce the
object $\theta_{\alpha,\beta}(z,\tau)$ ($\alpha,\beta=0,1/2$)\footnote{
  This object is almost identical to the one introduced in 
  ref.~\protect\cite{Izmailian4}. However, this latter one gives the wrong 
  sign to $\theta_1(z,\tau)$ [c.f.~\protect\reff{def_standard_thetas}], 
  although this is not important as we are only interested in the case $z=0$ 
  where $\theta_1(0,\tau)=0$.  
} 
\be
\label{def_theta_ab1}
\theta_{\alpha,\beta}(z,\tau) = \sum\limits_{n\in {\mathbb Z}} 
      q^{(n+1/2 -\alpha)^2} 
      \exp\left\{  2\pi i \left(n+{1\over 2}-\alpha  \right)
                          \left(z+\beta - {1\over 2} \right)
          \right\} 
\ee
where the nome $q$ is defined in terms of the modular parameter 
$\tau$ as follows
\be
\label{def_q}
q = e^{\pi i \tau} 
\ee
Using the identity (proved in \cite{Lawden}) 
\be
\prod\limits_{n=0}^\infty \left[ 1 + q^{2n-1} t      \right]
                          \left[ 1 + q^{2n-1} t^{-1} \right] 
                          \left[ 1 - q^{2n}          \right] = 
\sum\limits_{n\in {\mathbb Z}} q^{n^2} \, t^n 
\ee
we can write \reff{def_theta_ab1} as
\begin{eqnarray}
\theta_{\alpha,\beta}(z,\tau) &=& 
              \eta(\tau) \, q^{B_2(\alpha)}  
              e^{2 \pi i (1/2 -\alpha)(z +\beta-1/2)}  
                \nonumber \\
          & & \quad \times \prod\limits_{n=0}^\infty
              \left[ 1 - q^{2(n+1-\alpha)} e^{ 2\pi i(z+\beta)} \right]
              \left[ 1 - q^{2(n+  \alpha)} e^{-2\pi i(z+\beta)} \right]
\label{def_theta_ab2}
\end{eqnarray}
where $\eta(\tau)$ is Dedekind $\eta$-function 
\be
\label{def_eta}
\eta(\tau) = e^{\pi i \tau/12} \prod\limits_{n=1}^\infty 
             \left[ 1 - e^{2\pi i \tau n} \right] 
\ee
and $B_2(\alpha)$ is the Bernoulli polynomial [c.f.,\reff{def_Bna}] 
\be
\label{def_B2a}
B_2(\alpha) = \alpha^2 - \alpha + {1\over 6}
\ee

The relation of the functions $\theta_{\alpha,\beta}$ with the usual 
$\theta$-functions $\theta_i(z,\tau)$ $i=1,\ldots,4$ \cite{Whittaker} 
is the following
\begin{eqnarray}
\theta_{0,0}(z,\tau)                 &=& \theta_1(z,\tau) = 
  -i \sum\limits_{n\in {\mathbb Z}} 
 (-1)^n e^{ 2\pi i z(n+1/2) + \pi i \tau (n+1/2)^2} \\
\theta_{0,{1\over2}}(z,\tau)         &=& \theta_2(z,\tau) =  
 \phantom{-i} \sum\limits_{n\in {\mathbb Z}} 
 e^{ 2\pi i z(n+1/2) + \pi i \tau (n+1/2)^2}\\
\theta_{{1\over2},0}(z,\tau)         &=& \theta_4(z,\tau) =  
 \phantom{-i} \sum\limits_{n\in {\mathbb Z}} 
 (-1)^n e^{ 2\pi i z n + \pi i \tau n^2} \\
\theta_{{1\over2},{1\over2}}(z,\tau) &=& \theta_3(z,\tau) =  
 \phantom{-i} \sum\limits_{n\in {\mathbb Z}} 
 e^{ 2\pi i z n + \pi i \tau n^2} 
\label{def_standard_thetas}
\end{eqnarray} 

In this paper we will only need these functions evaluated at $z=0$ and 
$\tau = i \tau_0 \rho$ where $\tau_0$ is given by \reff{def_tau0}. To simplify 
the notation we will use the shorthands
\begin{subeqnarray}
\theta_{\alpha,\beta} 
         &=& \theta_{\alpha,\beta}(i\tau_0\rho) 
          = \theta_{\alpha,\beta}(z=0,\tau=i\tau_0\rho) \\  
\theta_i &=& \theta_i(i\tau_0\rho) = \theta_i(z=0,\tau=i\tau_0\rho) \\  
\eta     &=& \eta(i\tau_0\rho)     =         \eta(\tau=i\tau_0\rho) 
\label{shorthands_theta}
\end{subeqnarray}
We also need the limits of the $\theta$-functions in the limit 
$\rho\to\infty$. These limits are given by
\begin{subeqnarray}
\lim_{\rho\to\infty} \theta_3(i\tau_0\rho) &=& 
\lim_{\rho\to\infty} \theta_4(i\tau_0\rho) = 1 \\
\lim_{\rho\to\infty} \theta_2(i\tau_0\rho) &=& 
\lim_{\rho\to\infty} 2 e^{-\pi \tau_0 \rho/4} = 0 
\label{theta_limits}
\end{subeqnarray}

{}From eq.~\reff{def_theta_ab2} we arrive at the following identity
valid when $(\alpha,\beta)\neq (0,0)$: 
\begin{eqnarray}
\log \left| {\theta_{\alpha,\beta}(i\tau_0 \rho) 
       \over \eta(i\tau_0 \rho)} \right| + 
   \pi \rho \real(\tau_0) B_2(\alpha) &=& 
   \sum\limits_{n=0}^\infty \left\{  
  \log\left| 1 - e^{-2\pi [\tau_0 \rho (n+1-\alpha) - i \beta]}\right|
   \right. 
   \nonumber \\
   & &  
  \quad + \left.  
  \log\left| 1 - e^{-2\pi [\tau_0 \rho (n+\alpha) + i \beta]} \right|  
          \right\} 
\label{formula1}
\end{eqnarray}
Another useful relation involving $\log \theta_{\alpha,\beta}(0,\tau)$ is 
the following
\be
\sum\limits_{n=\delta_{\alpha,0}}^\infty \sum\limits_{p=1}^\infty 
{e^{2\pi p i(\tau(n+\alpha) - \beta)} \over n+\alpha} = - 
\left[ \log \theta_{\alpha,\beta}(\tau) - \left( {i \pi \tau\over 4} + \log 2 
     \right) \delta_{\alpha,0} \right] 
\label{relation_log_bis}
\ee 
We have proved this identity by considering each case $\alpha,\beta=0,1/2$ 
[with $(\alpha,\beta)\neq (0,0)$] separately and by a careful rearrangement of
the corresponding series.  

Dedekind's $\eta$-function satisfies the following identity 
\be
\label{formula3}
\eta(\tau)^3 = {1\over 2} \theta_2(\tau) \theta_3(\tau) \theta_4(\tau)
\ee
The analogue of \reff{formula1} when $(\alpha,\beta)=(0,0)$ is given in the 
particular case $\tau=i\tau_0\rho$ by  
\be
\label{formula2}
 \sum\limits_{n=1}^\infty \log \left| 1 - e^{-2\pi \tau_0 \rho n} \right| =
 \log|\eta| + {\pi \rho \over 12} \real(\tau_0)
\ee

We also need the behavior of the $\theta$ functions under the Jacobi 
transformation 
\be
\label{def_Jacobi_transformation}
\tau \rightarrow \tau' = -1/\tau 
\ee
The result when $z=0$ is given in ref.~\cite{Lawden} 
\begin{subeqnarray}
\theta_{3}(0,\tau')   &=& (-i\tau)^{1/2} \theta_{3}(0,\tau) \\
\theta_{2,4}(0,\tau') &=& (-i\tau)^{1/2} \theta_{4,2}(0,\tau) 
\label{def_Jacobi_transformation1}
\end{subeqnarray}
In particular, if $\tau=i\tau_0^\star\rho$ where
\be
\tau_0^\star = {\sqrt{3}+ i\over 2} = {1 \over \tau_0}  
\ee
is the complex conjugate of $\tau_0$ \reff{def_tau0}, the $\theta$-functions
transforms under \reff{def_Jacobi_transformation} as follows
\begin{subeqnarray}
\theta_{3}(0,i\tau_0^\star/\rho)   &=& (\tau_0\rho)^{1/2} 
                                        \theta_{3}(0,i\tau_0\rho) \\
\theta_{2,4}(0,i\tau_0^\star/\rho) &=& (\tau_0\rho)^{1/2} 
                                        \theta_{4,2}(0,i\tau_0\rho)
\label{def_Jacobi_transformation2}
\end{subeqnarray}
Finally, we should mention that the absolute value of the above 
$\theta$-functions does not depend on the sign of $\imag \tau_0$. Thus,
\be
|\theta_i(0,i\tau_0/\rho)| = |\theta_i(0,i\tau_0^\star/\rho)| 
\label{identity_abs_theta}
\ee
%

%
%
\section{Kronecker's double series} \label{sec_Kronecker}

In this appendix we collect a few properties of the Kronecker's double series
\cite{Weil}. These series are defined as
\be
\label{def_Kab}
K_p^{\alpha,\beta}(\tau) = - {p! \over (-2\pi i)^p }
  \mathop{\sum\limits_{m,n\in{\mathbb Z}}}_{(m,n)\neq (0,0)}  
  { e^{-2\pi i(n\alpha+m\beta)} \over (n+\tau m)^p }
\ee

The basic property we need is the following
\begin{eqnarray}
B_{2p}(\alpha) - \real K_{2p}^{\alpha,\beta}(\tau) &=& 
2p \real \sum\limits_{m=1}^\infty\sum\limits_{n=0}^\infty \left[ 
  (n+\alpha)^{2p-1} e^{2\pi i m[\tau(n+\alpha)-\beta]} \right. \nonumber \\
  & & \left. \qquad \quad + 
  (n+1-\alpha)^{2p-1} e^{2\pi i m[\tau(n+1-\alpha)-\beta]} \right]
\label{formulaK1}
\end{eqnarray} 
in the particular case $\tau=i\tau_0\rho$ with $\tau_0 \in {\mathbb C}$ 
[c.f., \reff{def_tau0}] and  $\rho\in{\mathbb R}$. 
Eq.~\reff{formulaK1} can be easily proved using the same arguments as in 
ref.~\cite[Appendix D]{Izmailian4} where they consider the particular 
case $\tau=i\rho$, $\rho\in{\mathbb R}$. 
  
In this paper we also need certain values of the $K_p^{\alpha,\beta}$
obtained in \cite[Appendix E]{Izmailian4}
\begin{subeqnarray}
\label{values_Kab_4}
K_4^{0,{1\over 2}}(\tau)
                &=& {1\over 30}
               \left({7\over 8}\theta_2^8 - \theta_3^4\theta_4^4\right) \\
K_4^{{1\over 2},0}(\tau)
                &=&{1\over 30}
               \left({7\over 8}\theta_4^8 - \theta_2^4\theta_3^4\right) \\
K_4^{{1\over 2},{1\over 2}}(\tau)
                &=& {1\over 30}
               \left({7\over 8}\theta_3^8 + \theta_2^4\theta_4^4\right)
\end{subeqnarray}
\begin{subeqnarray}
\label{values_Kab_6}
K_6^{0,0}(\tau) &=& \phantom{-} 
                    {1\over 84}(\theta_2^4 + \theta_3^4)
                               (\theta_4^4 - \theta_2^4)
                               (\theta_3^4 + \theta_4^4) \\
K_6^{0,{1\over 2}}(\tau) 
                &=& \phantom{-} 
                    {1\over 84}(\theta_3^4 + \theta_4^4)
               \left({31\over 16}\theta_2^8 + \theta_3^4\theta_4^4\right) \\
K_6^{{1\over 2},0}(\tau) 
                &=&-{1\over 84}(\theta_2^4 + \theta_3^4)
               \left({31\over 16}\theta_4^8 + \theta_2^4\theta_3^4\right) \\
K_6^{{1\over 2},{1\over 2}}(\tau) 
                &=& \phantom{-} 
                    {1\over 84}(\theta_2^4 - \theta_4^4)
               \left({31\over 16}\theta_3^8 - \theta_2^4\theta_4^4\right)
\end{subeqnarray}

The behavior of the functions $K_6^{\alpha,\beta}$ 
under the Jacobi transformation 
\reff{def_Jacobi_transformation} can be obtained using 
\reff{def_Jacobi_transformation2} and taking into account that 
$\tau_0^6 = -1$  
\begin{subeqnarray}
K_6^{0,{1\over2}}(0,i\tau_0^\star/\rho) &=& \rho^6 
                         K_6^{{1\over2},0}(0,i\tau_0\rho) \\
K_6^{{1\over2},0}(0,i\tau_0^\star/\rho) &=& \rho^6 
                         K_6^{0,{1\over2}}(0,i\tau_0\rho) \\ 
K_6^{{1\over2},{1\over 2}}(0,i\tau_0^\star/\rho) &=& \rho^6 
                         K_6^{{1\over 2},{1\over2}}(0,i\tau_0\rho) \\  
K_6^{0,0}(0,i\tau_0^\star/\rho) &=& \rho^6 
                         K_6^{0,0}(0,i\tau_0\rho)
\label{Jacobi_K6ab}
\end{subeqnarray}
Finally we mention that the value of $\real K_6^{\alpha,\beta}$ does not
depend on the sign of $\imag \tau_0$:  
\be
\real K_6^{\alpha,\beta}(0,i\tau_0^\star/\rho) = 
\real K_6^{\alpha,\beta}(0,i\tau_0/\rho)
\label{identity_re_K6ab}
\ee
%

%
%

\section*{Acknowledgments}

We thank Andrea Pelissetto and Alan Sokal for many useful comments and 
for a critical reading of the manuscript. We also thank Malte Henkel for
correspondence. 
The authors' research was supported in part by CICyT (Spain) grant
AEN99-0990.

%
%
\renewcommand{\baselinestretch}{1}
\large

\end{document}